\documentclass[twocolumn,aps,nofootinbib,longbibliography]{revtex4-2}

\usepackage[dvipdfmx]{graphicx}
\usepackage{mathtools,amssymb,amsmath,bm,url,color}

\usepackage[breaklinks]{hyperref}
\usepackage{multirow}
\allowdisplaybreaks[1]

\usepackage{mhchem} % \ce 

\newcommand{\argmax}{\mathop{\rm arg~max}}
\begin{document}

\title{
Hidden Topological Transitions in Emergent Magnetic Monopole Lattices
}

\author{Yasuyuki~Kato and Yukitoshi~Motome}

\affiliation{Department of Applied Physics, the University of Tokyo, Tokyo 113-8656, Japan}

\begin{abstract}
Topological defects, called magnetic hedgehogs, 
realize emergent magnetic monopoles, which are not allowed in the ordinary electromagnetism described by Maxwell's equations.
Such monopoles were experimentally discovered in magnets in two different forms: 
tetrahedral $4Q$ and cubic $3Q$ hedgehog lattices.
The spin textures are modulated by the chemical composition, 
an applied magnetic field, and temperature, 
leading to quantum transport and optical phenomena through movement and pair annihilation of magnetic monopoles, 
but the theoretical understanding remains elusive, especially in the regions where different types of hedgehog lattices are competing.
Here we propose a theoretical 
model that can stabilize both tetrahedral and cubic hedgehog lattices, 
and perform a thorough investigation of the phase diagram while changing the interaction parameters,
magnetic field, and temperature, by using a recently developed method that delivers exact solutions in the thermodynamic limit.
We find that the model exhibits various types of topological transitions with 
changes of the density of monopoles and antimonopoles, some of which are accompanied by
anomalies in the thermodynamic quantities, while the others are hidden with less or no anomaly. 
We also find another hidden topological transition with pair annihilation of two-dimensional vortices in the three-dimensional system.
These results not only provide useful information for understanding the existing experimental data, 
but also challenge the identification of hidden topological transitions and the exploration of emergent electromagnetism in magnetic monopole lattices.
\end{abstract}

\maketitle

%%%%%%%%%%%%%%%%%%%%%%%%%%%%%%%%%%%%
\section{Introduction}
%%%%%%%%%%%%%%%%%%%%%%%%%%%%%%%%%%%%
\begin{figure*}[!hbpt]
  \centering
  \includegraphics[trim=0 60 0 20, clip,width=\textwidth]{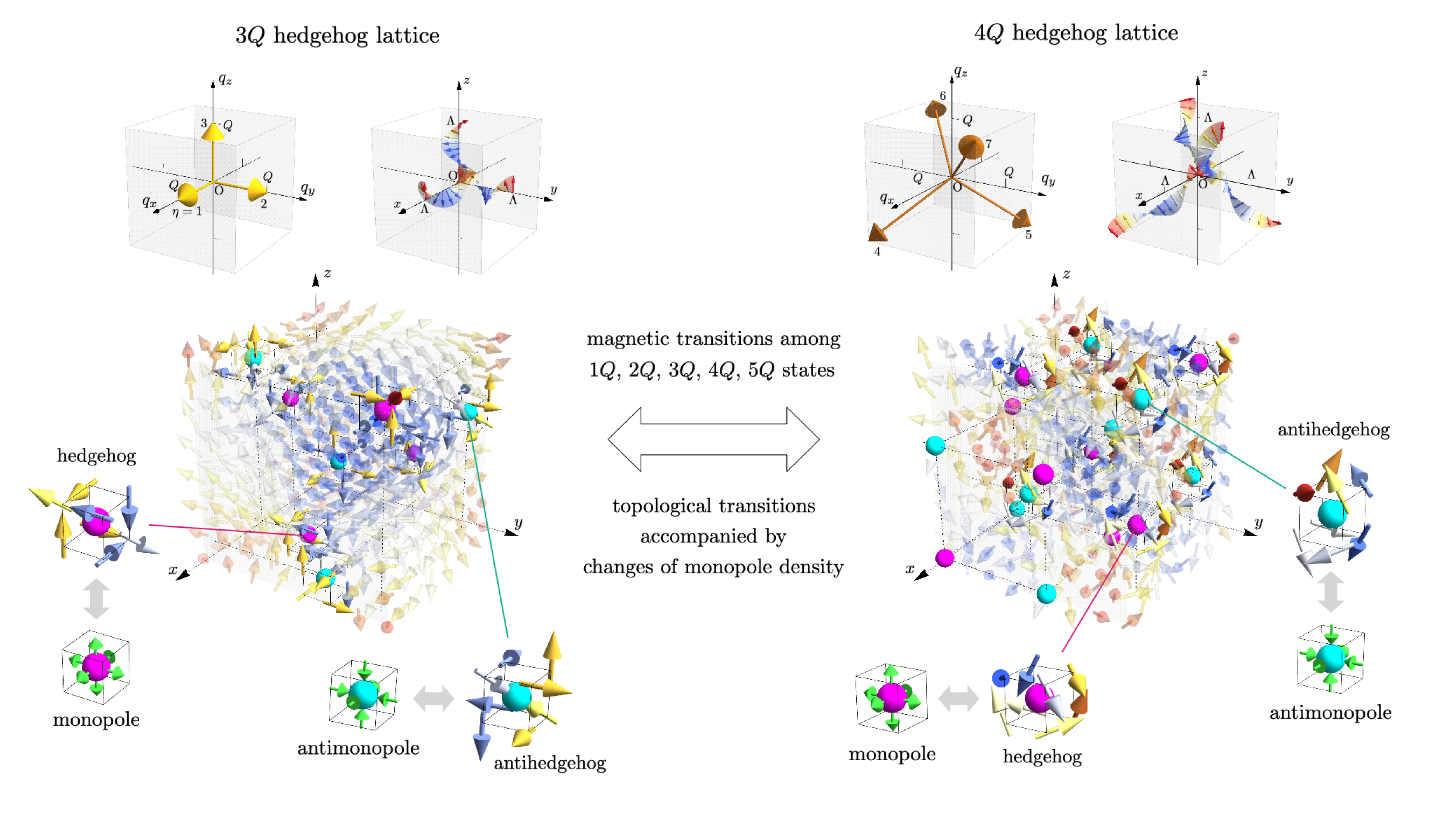}
  \caption{
  Cubic $3Q$ (left) and tetrahedral $4Q$ (right) HLs, and the possible transitions between them.
  Main panels show the position of hedgehogs and antihedgehogs in the $3Q$ and $4Q$ HLs with surrounding spins.
  Enlarged pictures of the spin configurations around the hedgehogs and antihedgehogs, 
  and the corresponding monopoles and antimonopoles defined by the sources and sinks of the emergent magnetic fields denoted by the light green arrows are also shown.
  Top panels show the characteristic wave numbers ${\bf Q}_\eta$ and the spin helices superposed to form the HLs.
  The labeled numbers indicate $\eta$, and $\Lambda$ denotes the period of the helices.
  The gray cubes are guides to the eye.
  }
  \label{fig01}
\end{figure*}
%%%%%%%%%%%%%%%%%%%%%%%%%%%%%%%%%%%%

Topological defects are found ubiquitously, from cosmology, meteorology, biology, and material science. 
In particular, in condensed matter physics, various types of the topological defects have been discovered, which play an important role in the properties of matter. 
Typical examples are found in magnets,
such as domain walls and vortices in swirling spin textures~\cite{Bogdanov1989,Braun2012,Seidel2016,Bogdanov2020}. 
Among such magnetic topological defects, magnetic hedgehog and antihedgehog are unique since they realize magnetic monopole and antimonopole, respectively~\cite{Milde2013}, 
which are not allowed in the ordinary electromagnetism described by Maxwell's equations 
--- these spin textures have spin vanishing singularities at the cores, which can be viewed as source and sink of an emergent magnetic flux quanta arising from the Berry phase mechanism~\cite{Volovik1987}. 
They are characterized by a topological index called monopole charge, which is defined by how many times the spins around the core wraps a unit sphere. 
The magnetic hedgehog and antihedgehog have attracted much interests since they are expected to generate unique electromagnetic phenomena beyond the ordinary electromagnetism and also carry information associated with their topological indices.

Such magnetic hedgehogs and antihedgehogs were experimentally discovered in \ce{MnSi_{1-$x$}Ge_{x}} in a form of a periodic lattice structure called the magnetic hedgehog lattice (HL)~\cite{Kanazawa2011,Kanazawa2012,Tanigaki2015,Kanazawa2016,Kanazawa2017,Tokura2017,Fujishiro2019,Kanazawa2020,Fujishiro2020}. 
Two different types of HLs have been reported depending on $x$: 
a tetrahedral HL for $0.3\lesssim x\lesssim 0.7$ and a cubic HL for $0.7\lesssim x \leq 1.0$. 
Both spin structures are regarded as a superposition of multiple spin helices. 
The tetrahedral HL is composed of four spin helices propagating along the directions from the center to four corners of a tetrahedron, and the cubic HL is composed of three spin helices propagating along the three orthogonal directions; hence, they are also called the $4Q$ and $3Q$ HLs (Fig.~\ref{fig01}). 
Similar tetrahedral $4Q$ HL was also found in \ce{SrFeO3}~\cite{Ishiwata2020}. 
In these HLs, unconventional quantum transport and optical phenomena, such as the topological Hall effect~\cite{Kanazawa2011,Kanazawa2012,Hayashi2021}, 
the topological Nernst effect~\cite{Shiomi2013}, 
and the magneto-Seebeck effect~\cite{Fujishiro2018}, were observed in an applied magnetic field, presumably associated with the unique electromagnetism of magnetic monopoles and antimonopoles.
Although further exotic phenomena could be expected in the competing region between the different types of HLs, no in-depth research has been reported thus far.

The emergent electromagnetic phenomena in HLs have also been studied theoretically. 
For instance, to explain the stability of HLs, different types of lattice spin models were proposed, 
by including short-range~\cite{Park2011,Yang2016},
long-range multiple spin interactions~\cite{Okumura2020,Okumura2020b,Shimizu2021a,Okumura2022},
and long-range anisotropic spin interactions~\cite{Kato2021,Kato2022}.
Effects of the magnetic field were also studied in detail, and interesting topological transitions were found to occur with pair annihilation of magnetic monopoles and antimonopoles~\cite{Okumura2020,Okumura2022}. 
However, comprehensive understanding has been unreached yet, especially including the competition between different types of HLs. 
The fundamental questions that we address in this study are (i) how different types of HLs transform into each other, 
(ii) how they respond to an external magnetic field, 
and (iii) how they behave at finite temperature. 
Although all these questions are crucial for not only understanding of the unique electromagnetism
of magnetic monopoles but also providing a guiding principle for further experimental exploration, 
these remain challenging issues because of the lack of appropriate models and 
the huge computational cost for the comprehensive study in three-dimensional systems.

In this paper, 
we propose a model that stabilizes two types of HLs, the $3Q$ and $4Q$ HLs, at zero magnetic field, and study their competing region while changing the magnetic field and temperature~(Fig.~\ref{fig01}).
By using the exact steepest descent method recently developed by the authors~\cite{Kato2022},
we clarify the phase diagram in the thermodynamic limit.
Through the analysis, we unveil a variety of the topological transitions with changes of the density of emergent magnetic monopoles and antimonopoles.
Notably, these topological transitions, when not accompanied by magnetic phase transitions, are ``hidden'', namely, they show less or no anomaly in the thermodynamic quantities, 
such as the specific heat and the magnetization.
We also find another type of a hidden topological transition not related to monopoles, 
caused by pair annihilation of two-dimensional vortices in the three-dimensional system.
It is worth noting that these hidden topological transitions are hard to detect by numerical studies such as the conventional brute-forced Monte Carlo simulation at finite temperature, 
and their complete identifications are captured for the first time by using the present exact method.

The structure of the paper is as follows.
In Secs.~\ref{sec:model} and \ref{sec:method}, we outline the theoretical model and the method 
for the analysis of the magnetic and topological transitions in the HLs.
In Sec.~\ref{sec:results}, we present the results of the ground-state phase diagrams including both $3Q$ and $4Q$ HLs,
and the magnetic field--temperature phase diagrams for three representative parameter sets in their competing region.
Finally, Sec.~\ref{sec:summary} is devoted for the summary.

%%%%%%%%%%%%%%%%%%%%%%%%%%%%%%%%%%%%
\section{Model}\label{sec:model}
%%%%%%%%%%%%%%%%%%%%%%%%%%%%%%%%%%%%

Since we are interested in the competition between different types of HLs 
as observed in MnSi$_{1-x}$Ge$_x$, we consider a model which can stabilize HLs even at zero magnetic field. 
A candidate is found in spin lattice models with long-range interactions mediated by itinerant electrons~\cite{Okumura2020,Okumura2022}, 
which are variants of the models studied for various types of swirling spin textures~\cite{Hayami2017,Hayami2018,Yasui2020,Yambe2021,Hirschberger2021,Hayami2021c,Hayami2021d,Hayami2021b,Hayami2021g,Shimizu2021a,Shimizu2021b,Kato2021,Khanh2022,Kato2022}. 
In the previous study~\cite{Okumura2020}, two models were independently studied for the cubic $3Q$ and tetrahedral $4Q$ HLs. 
To study the competition between the two,
we integrate the two models by interpolating the interaction parameters.
The Hamiltonian is given by
\begin{align}
\mathcal{H}=
2 \sum_\eta 
\Bigl[
&-J_\eta {\bf S}_{{\bf Q}_\eta} \cdot {\bf S}_{-{\bf Q}_\eta}
+\frac{K_\eta}{N} \bigl( {\bf S}_{{\bf Q}_\eta} \cdot {\bf S}_{-{\bf Q}_\eta} \bigr)^2 \nonumber\\
&- i {\bf D}_\eta \cdot {\bf S}_{{\bf Q}_\eta} \times {\bf S}_{-{\bf Q}_\eta}
\Bigr]
-\sum_{\bf r} {\bf h} \cdot {\bf S}_{\bf r},
\label{eq:Hamiltonian} 
\end{align}
where
\begin{align}
{\bf S}_{\bf Q} = \frac{1}{\sqrt{N}} \sum_{\bf r} {\bf S}_{\bf r} e^{- i {\bf Q} \cdot{\bf r}},
\end{align}
${\bf S}_{\bf r}=(S^x_{\bf r},S^y_{\bf r},S^z_{\bf r})$ denotes the spin degree of freedom at site ${\bf r}$ on a simple cubic lattice, 
and $N$ is the total number of spins; 
we consider the classical spin limit where ${\bf S}_{\bf r} \in \mathbb{R}^3$ and $|{\bf S}_{\bf r}| = 1$, for simplicity.
The first term of the Hamiltonian in Eq.~\eqref{eq:Hamiltonian} represents an effective long-range spin interaction of the Ruderman-Kittel-Kasuya-Yosida type~\cite{Ruderman1954,Kasuya1956,Yosida1957}, 
where ${\bf Q}_\eta$ with $\eta=1,2,\ldots ,7$ are the characteristic wave numbers given by the nesting vectors of the Fermi surfaces of itinerant electrons in the limit of weak spin-charge coupling~\cite{Hayami2017}.
The second term represents an effective biquadratic interaction, which is most dominant in the higher-order perturbation in terms of the spin-charge coupling~\cite{Hayami2017}. 
The third term describes an antisymmetric interaction of the Dzyaloshinskii-Moriya (DM) type~\cite{Dzyaloshinsky1958,Moriya1960}, where the DM vectors are taken parallel to the corresponding characteristic wave number as
\begin{align}
{\bf D}_\eta = D_\eta \frac{{\bf Q}_\eta}{|{\bf Q}_\eta|}.
\label{eq:DMvec}
\end{align}
The last term in Eq.~\eqref{eq:Hamiltonian} represents the Zeeman coupling with an external magnetic field ${\bf h}$.
For the three interaction terms, 
to describe both the cubic $3Q$ and tetrahedral $4Q$ HLs, we choose the characteristic wave numbers as 
\begin{align}
\label{eq:cubic_Q1-Q3}
&{\bf Q}_1 = \begin{pmatrix} +Q, 0, 0 \end{pmatrix},\ 
{\bf Q}_2 = \begin{pmatrix} 0, +Q, 0 \end{pmatrix},\ 
{\bf Q}_3 = \begin{pmatrix} 0, 0,+Q \end{pmatrix},\\
\label{eq:tetra_Q4-Q7}
&{\bf Q}_4 = \begin{pmatrix} +Q,-Q,-Q \end{pmatrix},\ 
{\bf Q}_5 = \begin{pmatrix} -Q,+Q,-Q \end{pmatrix},\nonumber\\
&{\bf Q}_6 = \begin{pmatrix} -Q,-Q,+Q \end{pmatrix},\ 
{\bf Q}_7 = \begin{pmatrix} +Q,+Q,+Q \end{pmatrix},
\end{align}
where the former three in Eq.~\eqref{eq:cubic_Q1-Q3} [the latter four in Eq.~\eqref{eq:tetra_Q4-Q7}] prefer the cubic $3Q$ (tetrahedral $4Q$) HL~\cite{Okumura2020}; see Fig.~\ref{fig01}. 
We parametrize the coupling constants to interpolate the cubic and tetrahedral cases:
\begin{align}
\begin{pmatrix} J_{\eta},K_{\eta}, D_{\eta} \end{pmatrix}=
\begin{pmatrix} J(1-p),&K(1-p),&D(1-p) \end{pmatrix}, 
\label{eq:JKDfor3Q}
\end{align}
for $\eta = 1,2,3$, and
\begin{align}
\begin{pmatrix} J_{\eta},K_{\eta}, D_{\eta} \end{pmatrix}=
\begin{pmatrix} Jp,&Kp,&Dp \end{pmatrix}, 
\label{eq:JKDfor4Q}
\end{align}
for $\eta = 4,5,6,7$, with the mixing ratio $0\leq p\leq 1$.
The model stabilizes the cubic $3Q$ (tetrahedral $4Q$) HL at $p=0$ ($p=1$) when both $K$ and $D$ are sufficiently large~\cite{Okumura2020,Shimizu2021a}. 
Thus, considering that the interaction parameters are derived from the Fermi surface nesting, the interpolation by $p$ implicitly assumes smooth deformation of the Fermi surface 
with switching of the nesting vectors between the cubic and tetrahedral types. 
In the following calculations, 
we take the energy unit as $J=1$ and the lattice constant as unity. 
We set $K=0.6$,
$D=0.3$, and $Q = 2\pi/\Lambda$ with $\Lambda=8$; 
$\Lambda$ corresponds to the magnetic period of the stable spin textures~\cite{Okumura2020}.

%%%%%%%%%%%%%%%%%%%%%%%%%%%%%%%%%%%%
\section{Method}\label{sec:method}
%%%%%%%%%%%%%%%%%%%%%%%%%%%%%%%%%%%%

Although the model in Eq.~\eqref{eq:Hamiltonian} is derived from the previous ones~\cite{Okumura2020}, 
the systematic study of the phase diagram while changing the mixing ratio $p$, the magnetic field ${\bf h}$, and temperature $T$ is not a simple task. 
Indeed, the previous studies were limited to only the ground state for $p=0$ and $1$ by using simulated annealing. 
In the present study, we adopt a steepest descent approach recently developed by the authors~\cite{Kato2022}.
This method can provide an exact solution for a class of models including Eq.~\eqref{eq:Hamiltonian} in the thermodynamic limit,
not only for the ground state but also at finite temperature. 
Although the details of the method is found in Ref.~\cite{Kato2022}, we briefly describe the framework below to make the present paper to be self-contained.

A key observation is that the Hamiltonian can be written in terms of the averaged spins for each sublattice, 
\begin{align}
\overline{\bf S}_{{\bf r}_0} = \frac{1}{N_{\rm MUC}} \sum_{\bf R} {\bf S}_{{\bf R}+{\bf r}_0},
\end{align}
where ${\bf R}$ and ${\bf r}_0$ are the position vectors of the magnetic unit cell (MUC) and the internal sublattice site, respectively. 
This is because, in the model in Eq.~\eqref{eq:Hamiltonian}, all the magnetically ordered states have the magnetic periods dictated by the wave numbers ${\bf Q}_\eta$ in Eqs.~\eqref{eq:cubic_Q1-Q3} and \eqref{eq:tetra_Q4-Q7}, and hence, the MUC always fits into a cube of $\Lambda^3(\equiv N_0)$ spins
with the translation vectors $(\Lambda,0,0)$, $(0,\Lambda,0)$, and $(0,0,\Lambda)$. 
Thus, ${\bf R}=\Lambda(N^x, N^y, N^z)$ 
and
${\bf r}_0=(r^x_0,r^y_0,r^z_0)$ with integers $N^\mu \in [0,L)$ and $r^\mu_0 \in [0, \Lambda)$;
$N_{\rm MUC} $ is the number of MUC, i.e.,
$N_{\rm MUC} = L^3$.
Then, one can write the partition function by using $\overline{\bf S}_{{\bf r}_0}$ as
\begin{align}
Z=\int \Bigl( \prod_{{\bf r}} d {\bf S}_{\bf r} \Bigr) e^{-\beta \mathcal{H} }
= \int \Bigr[
\prod_{{\bf r}_0} d\overline{\bf S}_{{\bf r}_0}
\rho_{L^d} ( \overline{\bf S}_{{\bf r}_0} )
\Bigr]
e^{-\beta \mathcal{H}},
\label{eq:Z}
\end{align}
where $\beta$ is the inverse temperature and $\rho_{L^d} ( \overline{\bf S}_{{\bf r}_0} )$ is the density of state for $\overline{\bf S}_{{\bf r}_0}$. 
Note that the dimension of integration in Eq.~\eqref{eq:Z} is reduced from $2 N_{\rm MUC} N_0$ to $3 N_0$. 
After some algebra using the Pearson random walk~\cite{Pearson1905,Kiefer1984}, we obtain
\begin{align}
Z\to\int \Bigl( \prod_{{\bf r}_0} d \overline{\bf S}_{{\bf r}_0} \Bigr) e^{N_{\rm MUC} G(\{ \overline{S}^\alpha_{{\bf r}_0} \} ) },
\label{eq:ZofG}
\end{align} 
where
\begin{align}
&G(\{ \overline{S}^\alpha_{{\bf r}_0} \} ) \nonumber\\
&\quad=
-\frac{\beta \mathcal{H}}{N_{\rm MUC}}+
\sum_{{\bf r}_0}\Bigl[
\ln \Bigl( \frac{4\pi \sinh v_{0{\bf r}_0} }{v_{0{\bf r}_0}} \Bigr) -v_{0{\bf r}_0} |\overline{\bf S}_{{\bf r}_0}|
\Bigr],\label{eq:G}
\end{align}
with
\begin{align}
|\overline{\bf S}_{{\bf r}_0}|=\coth v_{0{\bf r}_0} -\frac{1}{v_{0{\bf r}_0}} \equiv S ( v_{0{\bf r}_0}).
\label{eq:v0_S}
\end{align}
From Eq.~\eqref{eq:ZofG},
the partition function in the thermodynamic limit ($N_{\rm MUC}\to\infty$)
is obtained by using the steepest descent method as
\begin{align}
Z \sim e^{N_{\rm MUC} G(\{ \overline{ \overline{S}^\alpha_{{\bf r}_0} }\} ) },
\end{align}
with $\{ \overline{ \overline{S}^\alpha_{{\bf r}_0}} \}=\argmax_{ \{ \overline{S}^\alpha_{{\bf r}_0} \}  } G(\{ \overline{S}^\alpha_{{\bf r}_0} \} )$.
In other words, the solution in this method is obtained by maximizing $G(\{ \overline{S}^\alpha_{{\bf r}_0} \} )$. 
It is worth noting that Eq.~\eqref{eq:v0_S} is useful for the parametrization of $\overline{\bf S}_{{\bf r}_0}$ by enabling us to write
\begin{align}
\overline{\bf S}_{{\bf r}_0}=&
S ( v_{0{\bf r}_0})
\bigl(
 \cos \varphi_{{\bf r}_0} \sin \theta_{{\bf r}_0},
 \sin \varphi_{{\bf r}_0} \sin \theta_{{\bf r}_0},
 \cos \theta_{{\bf r}_0}
\bigr),
\end{align}
with three real numbers $v_{0{\bf r}_0}$, $\theta_{{\bf r}_0}$, and $\varphi_{{\bf r}_0}$. 
For the maximization of $G(\{ \overline{S}^\alpha_{{\bf r}_0} \} )$, 
we use NVIDIA A100 GPU with a JAX-based~\cite{jax2018github} library, Optax~\cite{optax2020github}.

From the optimized values of $\overline{ \overline{S}^\alpha_{{\bf r}_0} }$, 
the free energy per spin is obtained as
\begin{align}
f=-\frac{G(\{ \overline{ \overline{S}^\alpha_{{\bf r}_0} }\} )}{\beta N_0}.
\end{align}
In addition, the internal energy $\varepsilon$ and the specific heat $C$ per spin are computed as
\begin{align}
&\varepsilon = \frac{\langle \mathcal{H} \rangle}{N} = 
 \lim_{ \{  \overline{S}^\alpha_{{\bf r}_0} \} \to \{ \overline{ \overline{S}^\alpha_{{\bf r}_0} } \} } \frac{\mathcal{H}}{N},\\ 
&C=\frac{\partial{\varepsilon}}{\partial T},
\end{align}
respectively, where $\langle \cdots \rangle$ denotes the thermal average. 
As the real-space spin configuration is given by 
$\langle {\bf S}_{{\bf R}+{\bf r}_0} \rangle = \overline{ \overline{\bf S}_{{\bf r}_0}}$, 
the magnetization $m$ and the spin scalar chirality are computed as
\begin{align}
m=&\frac{1}{N_0 h} \sum_{{\bf r}_0} \overline{ \overline{\bf S}_{{\bf r}_0}} \cdot {\bf h},\\
\chi_{\rm sc}=&\frac{1}{N_0 h} \sum_{{\bf r}_0} {\boldsymbol \chi}_{{\bf r}_0} \cdot {\bf h}, 
\label{eq:chi_sc}
\end{align}
respectively, where $h =|{\bf h}|$; ${\boldsymbol \chi}_{{\bf r}_0}$ represents the local scalar spin chirality defined as~\cite{Okumura2020,Okumura2020b}
\begin{align}
\chi^\gamma_{{\bf r}_0} =&
\frac{1}{2} \sum_{\alpha,\beta,\nu_\alpha,\nu_\beta}
\varepsilon^{\alpha\beta\gamma} \nu_\alpha \nu_\beta 
\overline{ \overline{\bf S}_{{\bf r}_0}} \cdot
\bigl(
\overline{ \overline{\bf S}_{{\bf r}_0+\nu_\alpha \hat{\boldsymbol \alpha}}} \times
\overline{ \overline{\bf S}_{{\bf r}_0+\nu_\beta \hat{\boldsymbol \beta}}} 
\bigr),
\end{align}
where
$\alpha,\beta,\gamma=x,y,z$,
$\varepsilon^{\alpha\beta\gamma}$ is the Levi-Civita symbol,
$\nu_{\alpha(\beta)} = \pm 1$, and 
$\hat{\boldsymbol \alpha} (\hat{\boldsymbol \beta})$ is the unit translation vector in the $\alpha (\beta)$ direction.
To identify the magnetically ordered phases, we define the order parameter as
\begin{align}
m_\eta=\sqrt{ \frac{1}{N_0} \overline{ \overline{\bf S}_{{\bf Q}_\eta}} \cdot \overline{ \overline{\bf S}_{-{\bf Q}_\eta} } },
\label{eq:OP}
\end{align}
which corresponds to the square root of the normalized spin structure factor,
with
$\overline{ \overline{\bf S}_{{\bf Q}} }
= \frac{ 1}{\sqrt{N_0}} \sum_{{\bf r}_0}
\overline{ \overline{\bf S}_{{\bf r}_0}} e^{- i {\bf Q} \cdot {\bf r}_0}$.
Moreover, to investigate the topological property of the magnetically ordered phases, 
extending the previous studies~\cite{Okumura2020,Okumura2020b,Shimizu2021a,Kato2021,Okumura2022}, 
we identify the monopoles and antimonopoles by computing the monopole charge in each cubic unit of the cubic lattice from configurations of normalized spins 
$\langle {\bf S}_{{\bf R}+{\bf r}_0} \rangle/|\langle {\bf S}_{{\bf R}+{\bf r}_0} \rangle|$~\cite{Kato2022}. 
We also obtain the number of monopoles and antimonopoles per MUC, 
$N_{\rm m}$, to distinguish different topological phases.

%%%%%%%%%%%%%%%%%%%%%%%%%%%%%%%%%%%%
\section{Results}\label{sec:results}
%%%%%%%%%%%%%%%%%%%%%%%%%%%%%%%%%%%%

In this section, we show the results obtained for the model in Eq.~\eqref{eq:Hamiltonian} by using the steepest descent method described in the previous section.
In Sec.~\ref{sec:GS}, we present the $p$ dependences of the magnetic order parameters at zero magnetic field and zero temperature (Sec.~\ref{sec:GSzeroB}),
and the $p$--$h$ phase diagrams at zero temperature with 
three different magnetic field directions, ${\bf h}\parallel [100]$, $[110]$, and $[111]$ (Sec.~\ref{sec:GSpB}).
In Sec.~\ref{sec:FT}, we present the $h$--$T$ phase diagrams for the three directions of the magnetic field,
focusing on the competing regime between the $3Q$ and $4Q$ HLs, at $p=0.4$ (Sec.~\ref{sec:FTp0.4}), $p=0.5$ (Sec.~\ref{sec:FTp0.5}),
and $p=0.6$ (Sec.~\ref{sec:FTp0.6}).

%%%%%%%%%%%%%%%%%%%%%%%%%%%%%%%%%%%%
\subsection{Ground-state phase diagrams}\label{sec:GS}
%%%%%%%%%%%%%%%%%%%%%%%%%%%%%%%%%%%%

%%%%%%%%%%%%%%%%%%%%%%%%%%%%%%%%%%%%
\subsubsection{Zero magnetic field}\label{sec:GSzeroB}
%%%%%%%%%%%%%%%%%%%%%%%%%%%%%%%%%%%%
\begin{figure}[bhtp]
  \centering
  \includegraphics[trim=0 0 0 0, clip,width=\columnwidth]{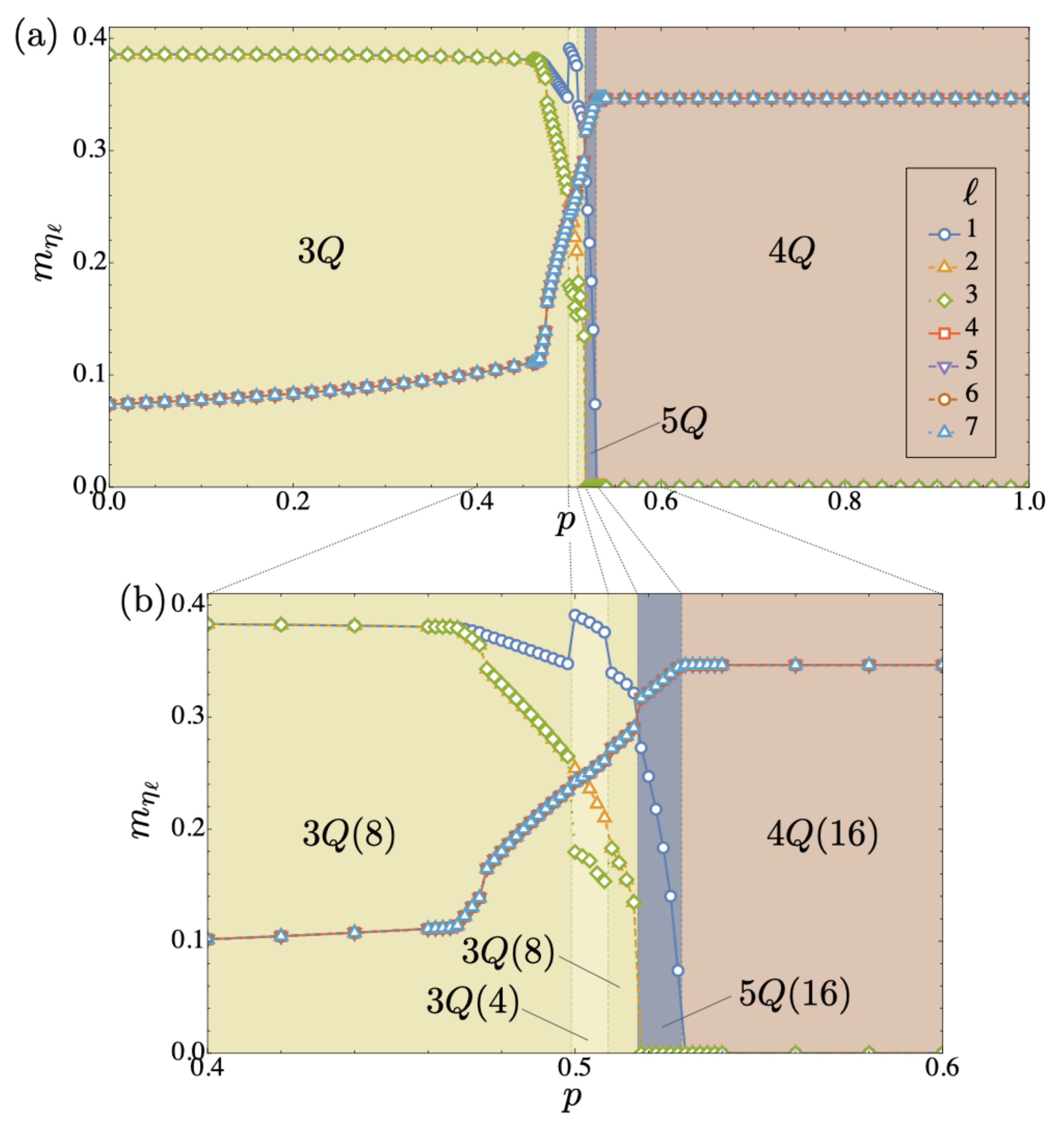}
  \caption{
  Order parameters $m_{\eta_\ell}$ at zero temperature as a function of the mixing ratio $p$ for the coupling constants in Eqs.~\eqref{eq:JKDfor3Q} and \eqref{eq:JKDfor4Q}:
  (a) the entire view for $0\leq p\leq 1$ and (b) an enlarged view for $0.4\leq p\leq 0.6$.
  $m_{\eta_\ell}$ are grouped into two as $\{ m_1, m_2, m_3 \}$ and $\{m_4,m_5,m_6,m_7\}$,
  and sorted in each group in descending order of the values of $m_{\eta_\ell}$. 
  The numbers in parentheses in (b) indicate the numbers of monopoles and antimonopoles per MUC, $N_{\rm m}$. 
  }
  \label{fig02}
\end{figure}
\begin{figure}[!hbtp]
  \centering
 \includegraphics[trim=0 0 0 0, clip,width=\columnwidth]{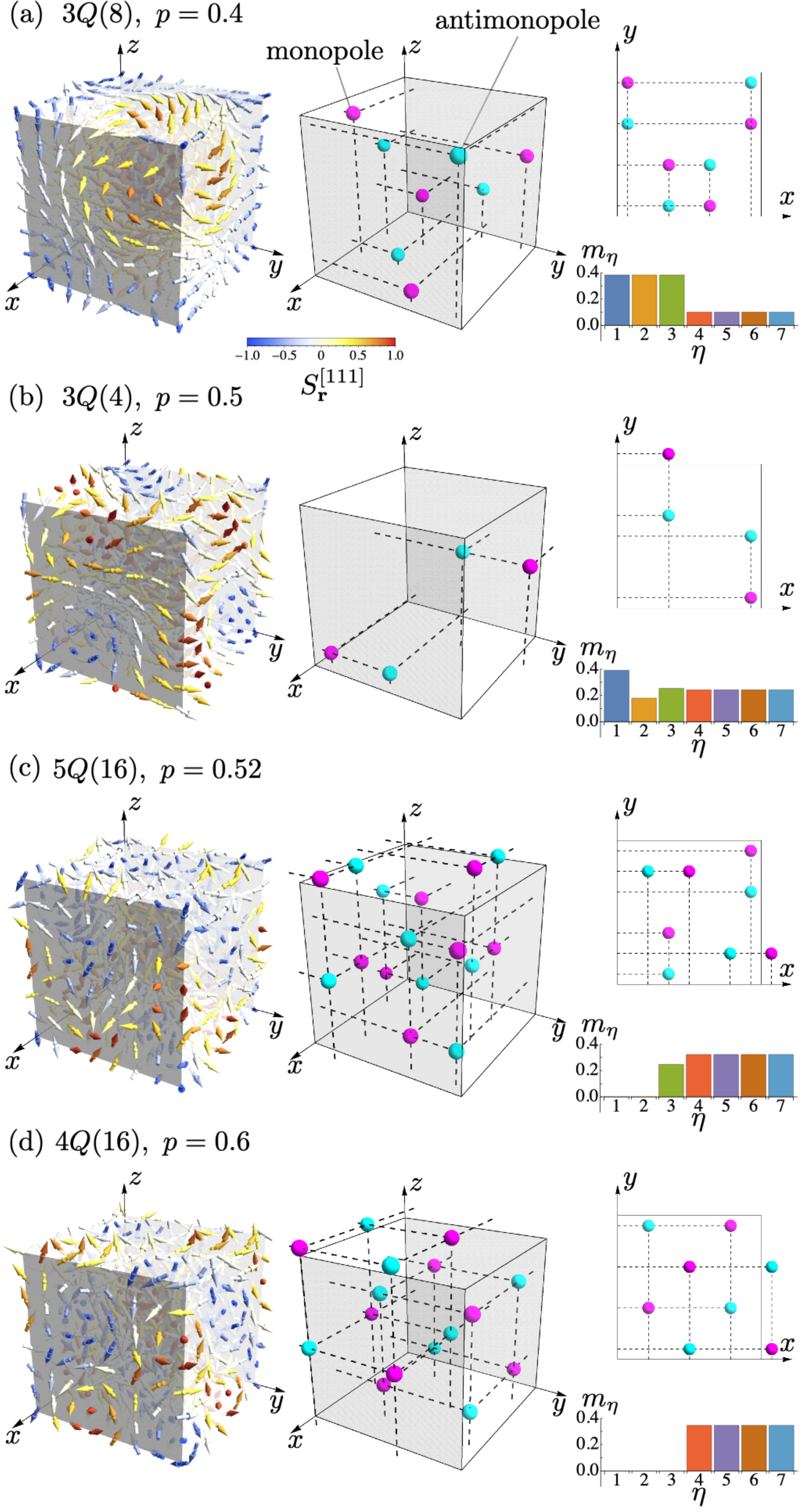}
  \caption{
  Ground-state spin configurations and positions of monopoles and antimonopoles for 
  (a) the $3Q$ state with $N_{\rm m}=8$ at $p=0.4$, 
  (b) the $3Q$ state with $N_{\rm m}=4$ at $p=0.5$, 
  (c) the $5Q$ state with $N_{\rm m}=16$ at $p=0.52$,  
  and (d) the $4Q$ state with $N_{\rm m}=16$ at $p=0.6$.
  The color of arrows represents the $[111]$ component of the spins, $S^{[111]}_{\bf r}$, as indicated in the inset of (a).
  The magenta and cyan spheres represent the monopoles (hedgehogs) and antimonopoles (antihedgehogs), respectively.
  The right panels show the top view of the HLs (top) and the distribution of the order parameters $m_\eta$ (bottom).
  }
  \label{fig03}
\end{figure}

\begin{table*}[!hbtp]
\caption{\label{tab1}
Magnetic phases, 
possible numbers of monopoles and antimonopoles per MUC $N_{\rm m}$, and conditions 
for the order parameters found in the ground-state and finite-temperature phase diagrams for the model in Eq.~\eqref{eq:Hamiltonian} with $J=1$, $K=0.6$, $D=0.3$, and $\Lambda = 8$.
The list of the related figures is also shown.
The starred $N_{\rm m}$ are found only at finite temperature. 
The possible sets of $\eta$ for nonzero $m_\eta$ are shown in the form of 
$\{ \eta_1, \eta_2, \cdots \}$ if necessary.
}
\begin{ruledtabular}
\begin{tabular}{lllll}
                   & Phase  & $N_{\rm m}$ & Conditions & Related figures\\
\hline
${\bf h}=0$ 
& $3Q$ & 0$^*$, 4, 8 & $m_{\eta}\neq 0$ for all $\eta = 1\text{--}3$. & \ref{fig02}, \ref{fig03}(a), \ref{fig03}(b), \ref{fig04}, \ref{fig06}, \ref{fig09}\\
& $4Q$ & 16 & $m_{\eta} =  0$ for all $\eta = 1\text{--}3$ with  $m_{\eta'}\neq 0$ for all $\eta' = 4\text{--}7$. & \ref{fig02}, \ref{fig03}(d), \ref{fig04}\\
& $5Q$ & 0$^*$, 16 & Five of $m_{\eta}$ are nonzero: $\{1,4\text{--}7\}$, $\{2,4\text{--}7\}$, and $\{3,4\text{--}7\}$. & \ref{fig02}, \ref{fig03}(c), \ref{fig09}
 \vspace{0.1cm}\\
\hline
${\bf h}\parallel [100]$ 
	& $1Q$  & 0 & One of $m_\eta$ is nonzero: $\{1\}$. & \ref{fig04}(a), \ref{fig05}(a), \ref{fig08}(a)\\
        & $3Q$ & 0, 4, 8  & (same as $3Q$ for ${\bf h}=0$)  & \ref{fig04}(a), \ref{fig05}(a), \ref{fig08}(a)\\ 
        & $3Q'$ & 0 & Three of $m_\eta$ are nonzero: $\{1,4,7\}$ and $\{1,5,6\}$. & \ref{fig04}(a), \ref{fig08}(a)\\
        & $4Q$ & 0, 8, 16 & (same as $4Q$ for ${\bf h}=0$).   & \ref{fig04}(a), \ref{fig11}(a)\\
        & $5Q$ & 0$^*$, 16 & Five of $m_{\eta}$ are nonzero:  $\{1,4\text{--}7\}$. & \ref{fig08}(a)
 \vspace{0.1cm}\\
\hline
${\bf h}\parallel [110]$ 
	& $1Q$  & 0 & One of $m_\eta$ is nonzero: $\{1\}$ and $\{2\}$. & \ref{fig04}(b), \ref{fig05}(b), \ref{fig07} \\
        & $1Q'$ & 0 & One of $m_\eta$ is nonzero: $\{6\}$ and $\{7\}$. & \ref{fig04}(b), \ref{fig08}(b), \ref{fig10}, \ref{fig11}(b) \\
        & $2Q$ & 0$^*$ & Two of $m_\eta$ are nonzero: $\{1,2\}$. & \ref{fig05}(b), \ref{fig07} \\
        & $2Q'$ & 0$^*$ & Two of $m_\eta$ are nonzero: $\{6,7\}$. & \ref{fig08}(b), \ref{fig11}(b) \\
        & $3Q$ & 0, 2, 4, 6, 8  &  (same as $3Q$ for ${\bf h}=0$) & \ref{fig04}(b), \ref{fig05}(b), \ref{fig07}, \ref{fig08}(b), \ref{fig10} \\
        & $3Q'$ & 0$^*$ & Three of $m_\eta$ are nonzero: $\{1,4,7\}$, $\{1,5,6\}$, $\{2,4,6\}$, and $\{2,5,7\}$.& \ref{fig08}(b)\\
        & $4Q$ & 0, 8, 16, 24$^*$ & (same as $4Q$ for ${\bf h}=0$) & \ref{fig04}(b), \ref{fig11}(b) \\
        & $5Q$ & 0$^*$, 4$^*$, 16 & Five of $m_{\eta}$ are nonzero: $\{1,4\text{--}7\}$ and $\{2,4\text{--}7\}$. & \ref{fig08}(b)
\vspace{0.1cm}\\
\hline
${\bf h}\parallel [111]$ 
	& $1Q$ & 0 & One of $m_\eta$ is nonzero: $\{1\}$, $\{2\}$, and $\{3\}$. & \ref{fig04}(c), \ref{fig05}(c) \\
        & $1Q'$ & 0 & One of $m_\eta$ is nonzero: $\{7\}$. & \ref{fig04}(c), \ref{fig08}(c), \ref{fig11}(c) \\
        & $3Q$ & 0, 2, 4, 6$^*$, 8  &  (same as $3Q$ for ${\bf h}=0$)& \ref{fig04}(c), \ref{fig05}(c), \ref{fig08}(c) \\
	& $3Q'$ & 0$^*$ & Three of $m_\eta$ are nonzero: $\{1,4,7\}$, $\{2,5,7\}$, and $\{3,6,7\}$. & \ref{fig08}(c) \\
        & $4Q$ & 8, 16 &  (same as $4Q$ for ${\bf h}=0$) &  \ref{fig04}(c), \ref{fig11}(c), \ref{fig12} \\
        & $5Q$ & 0$^*$ & (same as $5Q$ for ${\bf h}=0$) &  \ref{fig08}(c) \\
%%%%
\end{tabular}
\end{ruledtabular}
\end{table*}
%%%%%%%%%%%%%%%%%%%%%%%%%%%%%%%%%%%%

Figure~\ref{fig02} shows the mixing ratio $p$ dependences of the order parameters $m_{\eta_\ell}$~[Eq.~\eqref{eq:OP}] for the ground state.
The label of wave numbers $\eta$ are grouped into $1$--$3$ and $4$--$7$, 
sorted within each group, and assigned $\ell$ in descending order as 
$m_{\eta_1} \geq m_{\eta_2} \geq m_{\eta_3}$ and
$m_{\eta_4} \geq m_{\eta_5} \geq m_{\eta_6}\geq m_{\eta_7}$.

With an increase of $p$,
the ground state changes from the $3Q$ phase to the $5Q$ phase, and then to the $4Q$ phase, as shown in Fig.~\ref{fig02}(a).
In the $3Q$ phase, not only $m_{1\text{--}3}$ but also $m_{4\text{--}7}$ are nonzero, 
since $m_{4\text{--}7}$ correspond to higher harmonics of $m_{1\text{--}3}$, e.g., ${\bf Q}_4={\bf Q}_1-{\bf Q}_2-{\bf Q}_3$. 
Meanwhile, in the $4Q$ phase, $m_{1\text{--}3}=0$, and only $m_{4\text{--}7}$ are nonzero.
In the intermediate $5Q$ phase, one of the $3Q$ components, $m_1$, 
remains nonzero in addition to all the $4Q$ components $m_{4\text{--}7}$.
As shown in the enlarged view in Fig.~\ref{fig02}(b), the $3Q$--$5Q$ transition at $p \simeq 0.517$ is discontinuous 
with jumps of the order parameters, 
while the $5Q$--$4Q$ transition at $p \simeq 0.529$ is continuous.

In the entire range of $p$, the ground-state spin configuration contains topological defects, monopoles and antimonopoles, indicating that all the $3Q$, $4Q$, and $5Q$ states are HLs.
The number of monopoles and antimonopoles per MUC, $N_{\rm m}$, is indicated in the parentheses in Fig.~\ref{fig02}(b), 
and the ground-state spin configurations, 
the positions of the monopole and antimonopoles, and the distributions of $m_\eta$ for representative parameters at $p=0.4$, $0.5$, $0.52$, and $0.6$ are shown in Fig.~\ref{fig03}. 
A closer look reveals additional phase transitions within the $3Q$ phase: 
two discontinuous transitions at $p \simeq 0.499$ and $p\simeq 0.509$, 
where $N_{\rm m}$ is reduced from $8$ to $4$ and recovered from $4$ to $8$, respectively, as shown in Fig.~\ref{fig02}(b). 
This means that an intermediate $3Q$ phase with $N_{\rm m}=4$ intervenes in the $3Q$ phase with $N_{\rm m}=8$. 
At these discontinuous transitions, $m_{\eta_l}$ also changes discontinuously: $m_{\eta_2} = m_{\eta_3}$ for the $N_{\rm m}=8$ phase, 
while $m_{\eta_2}>m_{\eta_3}$ for the $N_{\rm m}=4$ phase. 
Thus, these transitions are magnetic phase transitions with topological changes. 
In addition, we find phase transitions within the $N_{\rm m}=8$ state:
a continuous one from $m_{\eta_1}=m_{\eta_2}=m_{\eta_3}$ to $m_{\eta_1}>m_{\eta_2}=m_{\eta_3}$ at $p\simeq 0.467$
and a discontinuous  one at $p\simeq 0.475$. 
We do not discuss these transitions in this paper because they are not relevant in the topological point of view.
All the phases with different $N_{\rm m}$ are summarized in the top row for ${\bf h}=0$ of Table~\ref{tab1}.

Let us make two remarks. 
One is on the competition between the $3Q$ and $4Q$ phases. 
In Fig.~\ref{fig02}, the $3Q$ region is slightly wider than the $4Q$ one. 
This is presumably because the characteristic wave numbers of the $4Q$ components coincide with the higher harmonics of those of the $3Q$ components in the present model, which may work in favor of stabilizing the $3Q$ phase over $4Q$. 
The other remark is on the intermediate $5Q$ phase. 
This is a peculiar phase with spontaneous symmetry breaking by selecting one of the $3Q$ wave numbers, which has never been reported in experiments to our knowledge. 
While further theoretical studies are necessary on the stability of this phase in a wider parameter range, 
our result would stimulate the experimental exploration in the competing regime between the $3Q$ and $4Q$ HLs.

%%%%%%%%%%%%%%%%%%%%%%%%%%%%%%%%%%%%
\subsubsection{Magnetic field dependence}\label{sec:GSpB}
%%%%%%%%%%%%%%%%%%%%%%%%%%%%%%%%%%%%
\begin{figure}[!htp]
  \centering
  \includegraphics[trim=0 0 0 0, clip,width=\columnwidth]{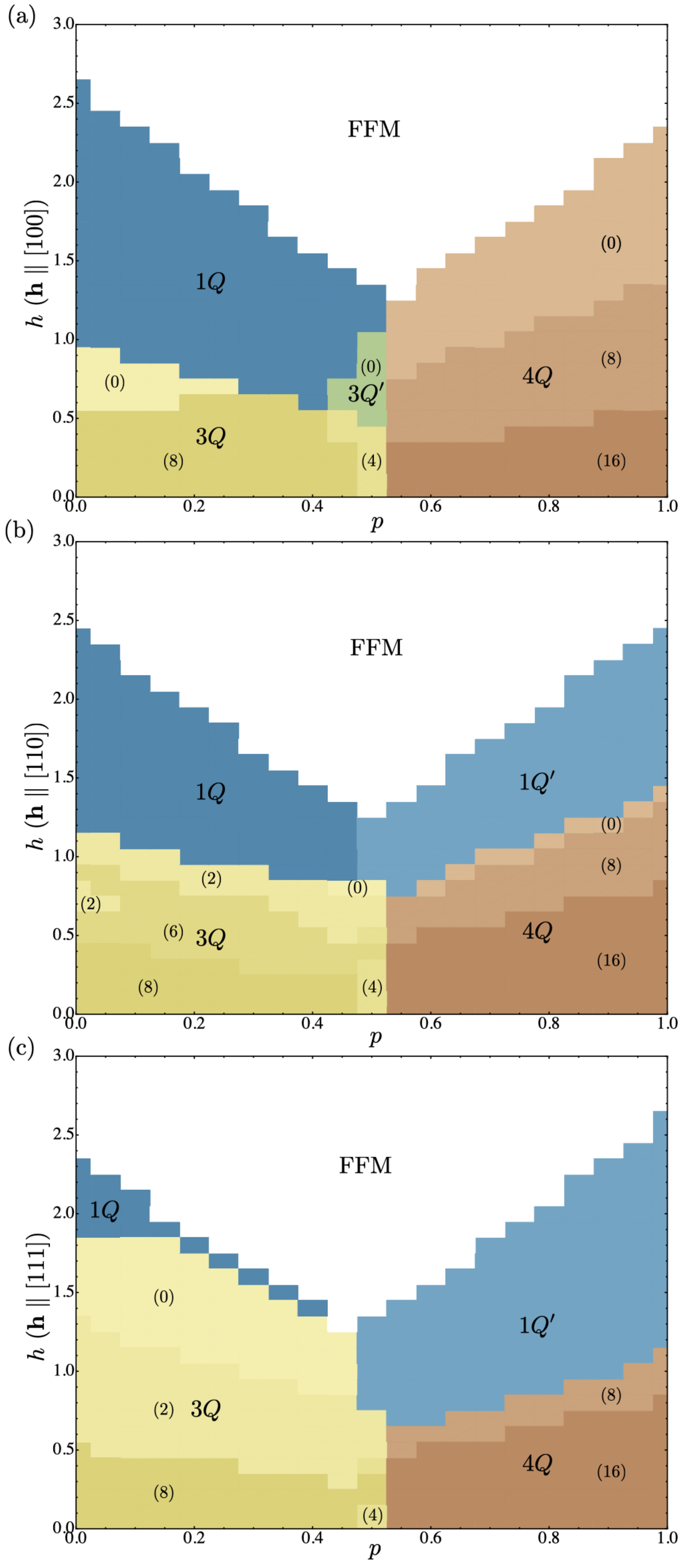}
  \caption{
  Ground-state phase diagrams on the $p$--$h$ plane for (a) ${\bf h}\parallel [100]$, (b) ${\bf h}\parallel [110]$, and (c) ${\bf h}\parallel [111]$.
  The calculations are done for every $0.05$ ($0.1$) with respect to $p$ ($h$).
  The colored regimes indicate the magnetically ordered phases while the white regimes indicate the forced ferromagnetic (FFM) phase.
  The numbers in parentheses represent the number of monopoles in a MUC, $N_{\rm m}$.
  }
  \label{fig04}
\end{figure}

Figure~\ref{fig04} shows the $p$--$h$ phase diagram for the cases of ${\bf h} \parallel [100]$, ${\bf h}\parallel [110]$, and ${\bf h}\parallel [111]$.
At $h=0$, only the $3Q$ and $4Q$ phases appear 
and the intermediate $5Q$ phase found in the previous section is absent because of the low resolution in the parameter setting; 
here, the calculations are done for every $0.05$ with respect to $p$.
When the magnetic field is applied, both $3Q$ and $4Q$ phases remain stable, but in most cases, 
they turn into a single-$Q$ phase ($1Q$ or $1Q'$) before the saturation to the forced ferromagnetic (FFM) phase, 
which is connected to the paramagnetic (PM) state at finite temperature.
The stability of the $3Q$ and $4Q$ phases depends on the relative directions of ${\bf h}$ and ${\bf Q}_\eta$: 
When ${\bf h}$ is applied in parallel to one of ${\bf Q}_\eta$, the $3Q$ and $4Q$ phases turn into the single-$Q$ phase characterized by ${\bf Q}_\eta \parallel {\bf h}$ at a relatively weak magnetic field, 
and the single-$Q$ phase is stable in a relatively wider field range; 
see Fig.~\ref{fig04}(a) for the $3Q$ case where ${\bf Q}_1 \parallel {\bf h}$ and Fig.~\ref{fig04}(c) for the $4Q$ case where ${\bf Q}_7 \parallel {\bf h}$. 
In the other field directions, the $3Q$ and $4Q$ phases remain more stable, and the single-$Q$ phases become narrower; 
the extreme case is the $4Q$ case under ${\bf h}\parallel [100]$, which shows direct saturation to FFM without any single-$Q$ phase.

It is worth noting that the relative stability between the $3Q$ and $4Q$ states does not change so much under the magnetic field,
namely, the phase boundaries between them are almost independent of $h$ within the present resolution. 
This is similar to the experimental results for \ce{MnSi_{1-$x$}Ge_{x}}~\cite{Fujishiro2019}.
In the competing region near $p=0.5$ in $h\parallel [100]$, however, we find complex successive phase transitions from $3Q$ to $3Q'$ and to $1Q$ before the saturation.
The $3Q'$ phase is found only in the finite field region, and the magnetic order is composed of a mixture of the $3Q$ and $4Q$ components  similar to the $5Q$ phase discussed in the previous section: 
two out of the $4Q$ components $\{ m_4 ,m_7\}$ or $\{m_5,m_6\}$ and one of the $3Q$ component $m_1$. 
Interestingly, the spin texture in the $3Q'$ phase exhibits two-dimensional modulation on the [$\bar{1}\bar{1}1$] or [$\bar{1}11$] plane since the three wave numbers are coplanar.
We will discuss an interesting field direction dependence of the spin scalar chirality in this two-dimensional $3Q'$ phase in Sec.~\ref{sec:FTp0.5}.

Both $3Q$ and $4Q$ phases exhibit various types of topological transitions accompanied by changes of $N_{\rm m}$ under the magnetic field.
In the $4Q$ case, $N_{\rm m}$ decreases monotonically, from $16$ to $8$ for all the field directions, and then 
from $8$ to $0$ before the transition to $1Q'$ or the saturation to FFM for the cases of ${\bf h}\parallel [100]$ and $[110]$.
In contrast, in the $3Q$ case, 
$N_{\rm m}$ does not always decrease monotonically and shows more complex field dependence.
For example, in the case of ${\bf h}\parallel [110]$,
$N_{\rm m}$ changes as $8\to 6 \to 2 \to 6 \to 2$ before the transition to the $1Q$ phase, as shown in Fig.~\ref{fig04}(b). 
We will study how these topological transitions evolve with temperature and how thermodynamic quantities behave at these transitions in the next section.

%%%%%%%%%%%%%%%%%%%%%%%%%%%%%%%%%%%%
\subsection{Magnetic field--temperature phase diagrams}\label{sec:FT}
%%%%%%%%%%%%%%%%%%%%%%%%%%%%%%%%%%%%

In this section, we present the results of the magnetic field--temperature phase diagrams. 
Besides magnetic phase transitions, we discuss topological transitions associated with changes of $N_{\rm m}$ caused by varying the temperature and the magnetic field. 
Focusing on the competing region between the $3Q$ and $4Q$ phases, 
we show the results for three values of $p$: $p=0.4$, $0.5$, and $0.6$ in Secs.~\ref{sec:FTp0.4}, \ref{sec:FTp0.5}, and \ref{sec:FTp0.6}, respectively.

%%%%%%%%%%%%%%%%%%%%%%%%%%%%%%%%%%%%
\subsubsection{$p=0.4$}\label{sec:FTp0.4}
%%%%%%%%%%%%%%%%%%%%%%%%%%%%%%%%%%%%
\begin{figure}[!htp]
  \centering
  \includegraphics[trim=0 10 0 0, clip,width=\columnwidth]{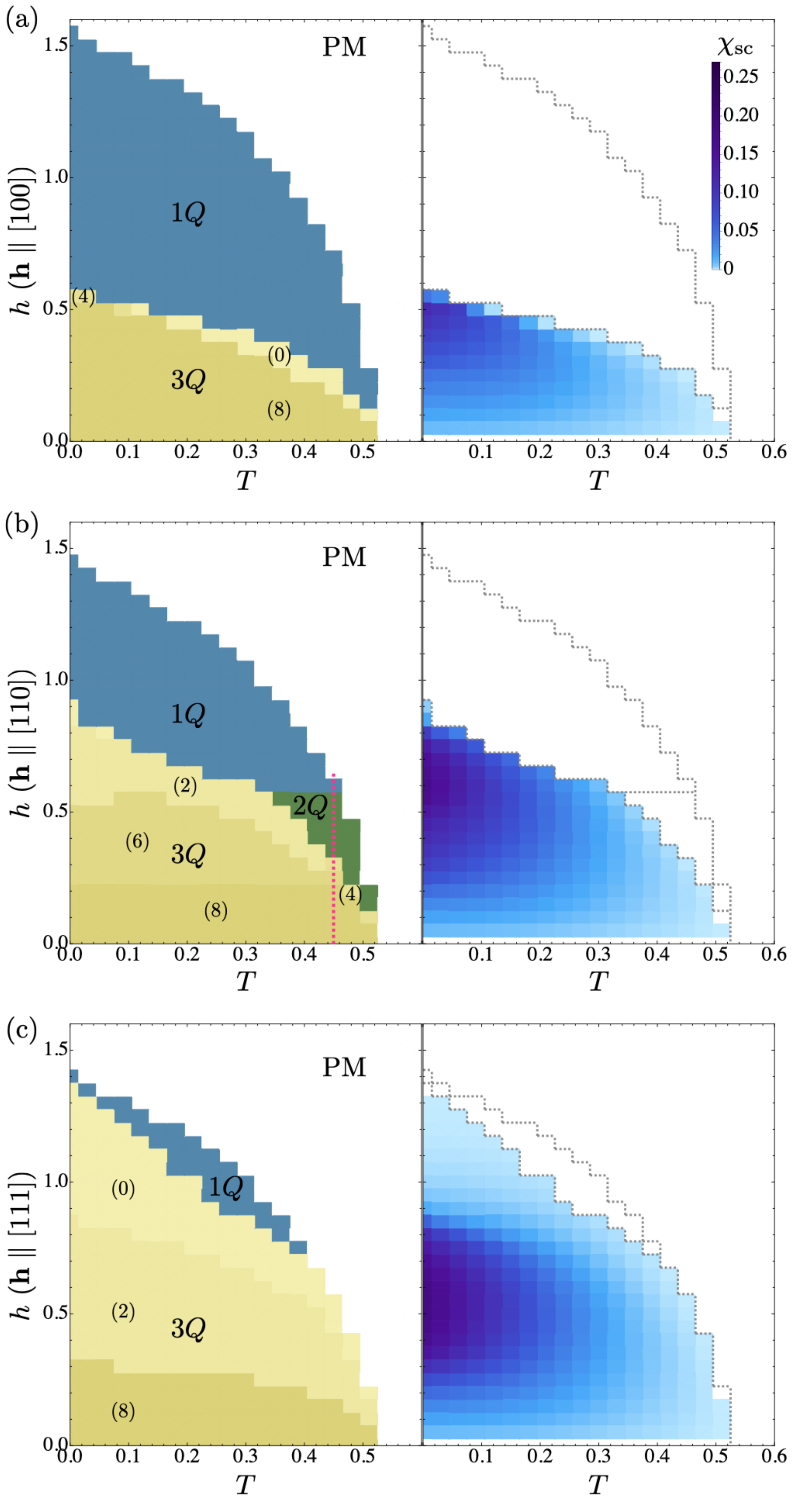}
  \caption{
  Magnetic field--temperature phase diagrams (left)
  and  $h$--$T$ maps of the spin scalar chirality $\chi_{{\rm sc}}$ (right) for 
  (a) ${\bf h}\parallel [100]$,
  (b) ${\bf h}\parallel [110]$, and
  (c) ${\bf h}\parallel [111]$
  at $p=0.4$.
  The calculations are done for every $0.05$ ($0.02$) with respect to $h$ ($T$). 
  The magenta dashed line in the left panel of (b) indicates the parameter range for Fig.~\ref{fig07}. 
  The gray dashed lines in the right panels represent the phase boundaries for guides to the eye. 
  }
  \label{fig05}
\end{figure}
%%%%%%%%%%%%%%%%%%%%%%%%%%%%%%%%%%%%
\begin{figure}[!htp]
  \centering
  \includegraphics[trim=0 20 0 0, clip,width=\columnwidth]{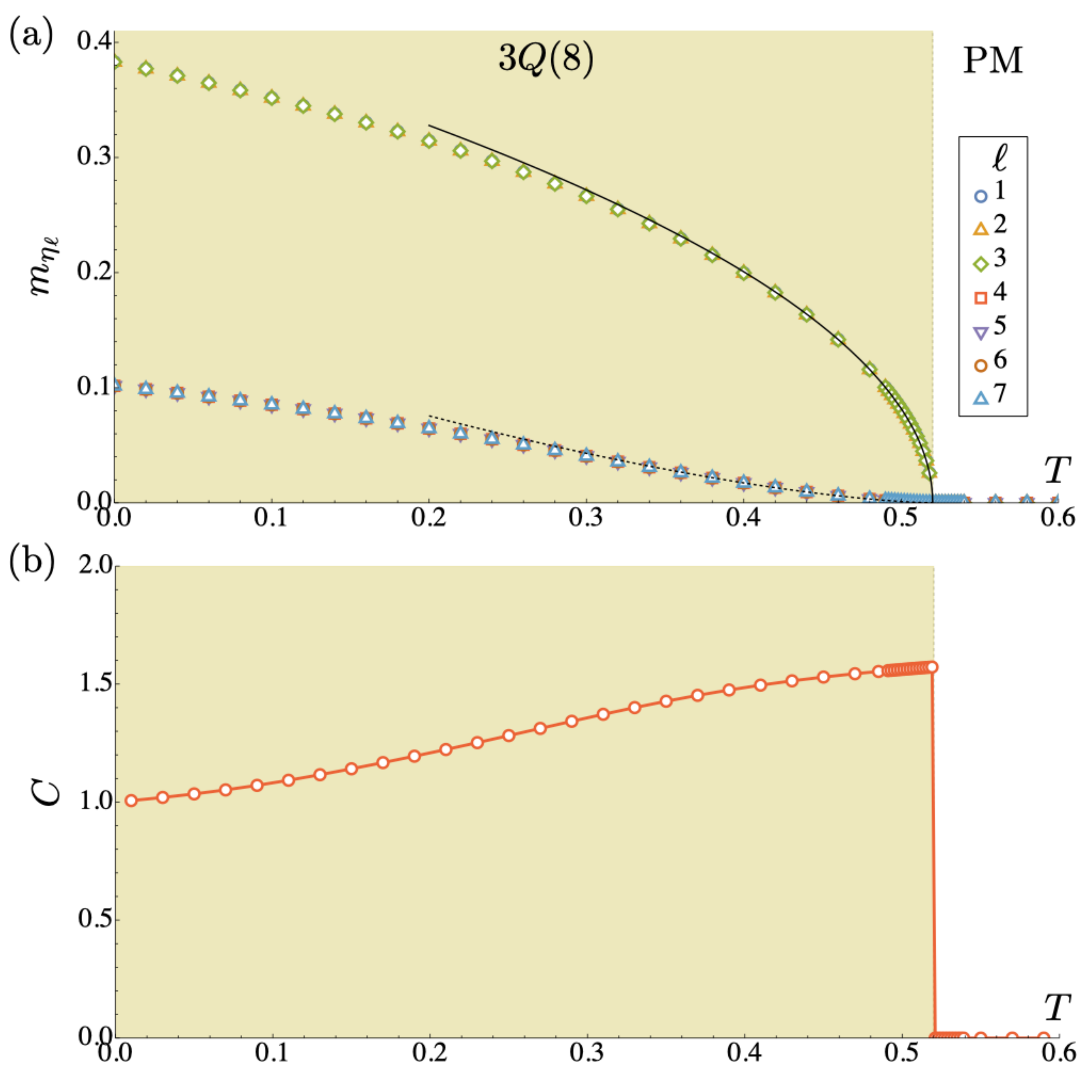}
  \caption{
  Temperature dependences of (a) the order parameters $m_{\eta_\ell}$ and (b) the specific heat $C$ at $h=0$ and $p=0.4$.
  The black solid and dashed lines in (a) indicate functions proportional to $|T-T_c|^{1/2}$ and $|T-T_c|^{3/2}$, respectively.
  }
  \label{fig06}
\end{figure}
%%%%%%%%%%%%%%%%%%%%%%%%%%%%%%%%%%%%

Let us begin with the case of $p=0.4$, which is close to the competing region but still in the $3Q$ phase; see Figs.~\ref{fig02} and \ref{fig04}. 
Figure~\ref{fig05} shows the $h$--$T$ phase diagrams at $p=0.4$ for three different directions of the magnetic field, 
${\bf h}\parallel [100]$, $[110]$, and $[111]$ (left panels).
At $h=0$, the ground state is the $3Q$ HL with $N_{\rm m}=8$, namely, four monopole--antimonopole pairs within the MUC; see Fig.~\ref{fig03}(a).
The $3Q$ phase remains stable against raising temperature, 
and shows a phase transition to the PM phase at $T\simeq 0.52$.
Figure~\ref{fig06} shows the temperature dependences of the order parameters $m_\eta$ and the specific heat per spin $C$ at $h=0$.
Near the critical temperature $T_c$, $m_{1\text{--}3}$ exhibit the critical behavior of the mean-field universality class as $m_{1\text{--}3} \propto |T-T_c|^{1/2}$, 
while $m_{4\text{--}7}$ behave differently as $m_{4\text{--}7} \propto |T-T_c|^{3/2}$. 
This indicates 
that $m_{1\text{--}3}$ are the primary order parameters, 
and $m_{4\text{--}7}$, which correspond to higher harmonics of $m_{1\text{--}3}$
as discussed in Sec.~\ref{sec:GSzeroB}, are induced as the secondary order parameters.
We note that the mean-field universality class is consistently understood from the fact that the range of the interactions in Eq.~\eqref{eq:Hamiltonian} is infinite.
We find that $C$ shows a jump at $T_c$ and becomes zero for $T>T_c$, which is also consistent with the mean-field universality class.

%%%%%%%%%%%%%%%%%%%%%%%%%%%%%%%%%%%%
\begin{figure}[bhtp]
  \centering
  \includegraphics[trim=0 50 0 0, clip,width=\columnwidth]{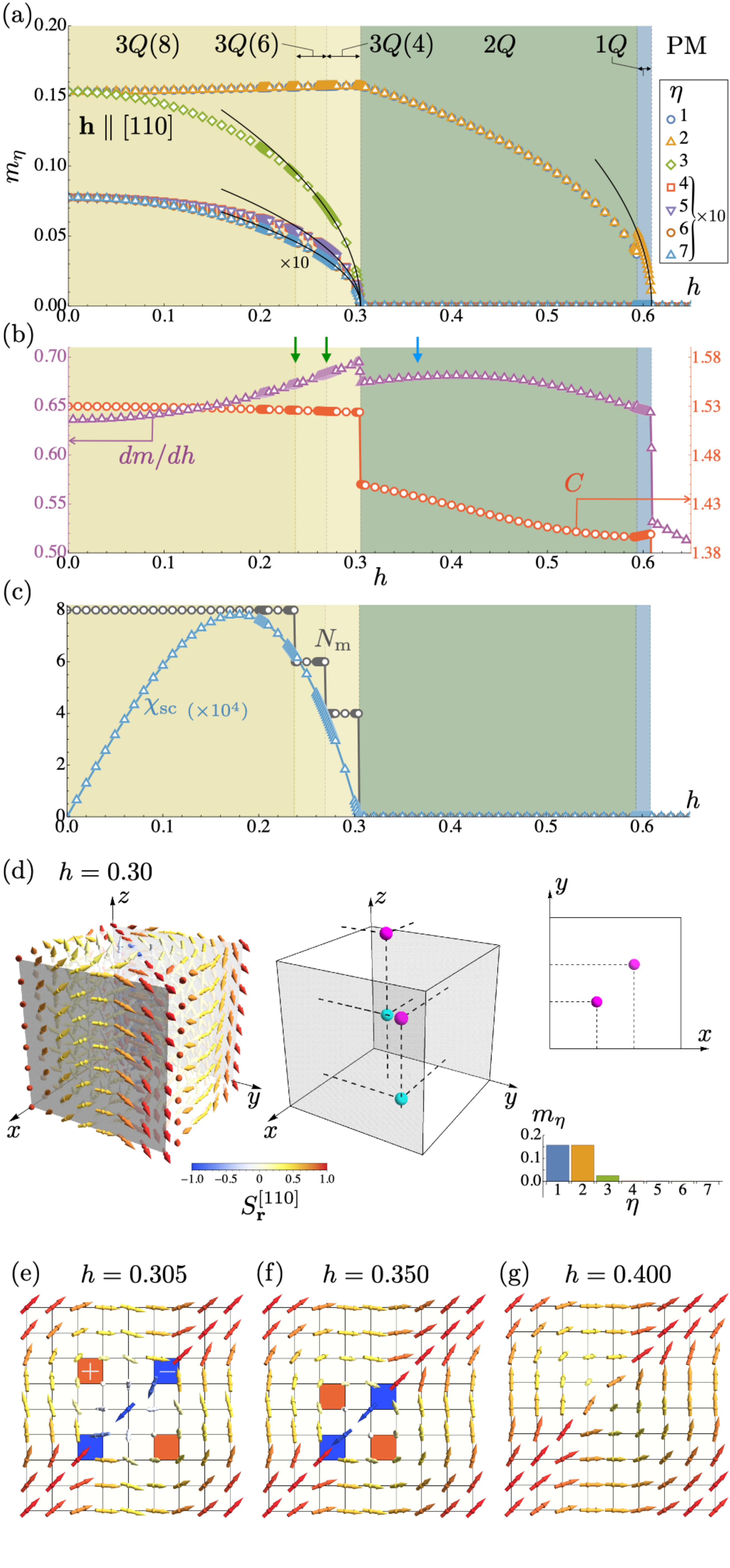}
  \caption{
  Magnetic field dependences of (a) $m_\eta$, 
  (b) $C$ and $dm/dh$, and (c) $N_{\rm m}$ and $\chi_{\rm sc}$. 
  The calculations are done for ${\bf h}\parallel [110]$ at $T=0.45$ and $p=0.4$; see the magenta dashed line in the left panel of Fig.~\ref{fig05}(b). 
  The data for $m_{4\text{--}7}$ in (a) and $\chi_{\rm sc}$ in (c) are multiplied by factors of $10$ and $10^4$, respectively, for better visibility.
  The solid lines in (a) indicate functions proportional to $|T-T_c|^{1/2}$. 
  The green and blue arrows in (b) indicate the hidden topological transitions associated with monopoles and vortices, respectively.
  (d) Spin configurations and positions of monopoles and antimonopoles at $h=0.3$.
  The inset shows distribution of $m_\eta$.
  Two-dimensional spin configurations on a $(001)$ slice at (e) $h=0.305$, 
  (f) $h=0.350$, and (g) $h=0.400$ in the $2Q$ phase.
  The orange and blue plaquettes represent the cores of vortices and antivortices.
  }
  \label{fig07}
\end{figure}
%%%%%%%%%%%%%%%%%%%%%%%%%%%%%%%%%%%%

In an applied magnetic field, the $3Q$ phase remains stable with changes of $N_{\rm m}$, but 
it undergoes phase transitions to the $1Q$ phase and then to the PM phase at low temperature, as seen in the ground state in Sec.~\ref{sec:GSpB}.
In contrast, we find different behaviors in the high-temperature region; 
the system undergoes a phase transition to the $2Q$ phase before entering into PM in ${\bf h}\parallel [110]$ [Fig.~\ref{fig05}(b)] and 
a direct transition from the $3Q$ phase to PM in ${\bf h}\parallel [111]$ [Fig.~\ref{fig05}(c)]. 
In addition, within the $3Q$ phase we find topological transitions that are not seen in the low-temperature region: 
the transition from $N_{\rm m}=8$ to $0$ in ${\bf h}\parallel [100]$ [Fig.~\ref{fig05}(a)] and $N_{\rm m}=6$ to $4$ in ${\bf h}\parallel [110]$ [Fig.~\ref{fig05}(b)]. 
As shown in the right panels of Fig.~\ref{fig05}, 
$\chi_{\rm sc}$ has nonzero values in the entire regions of the $3Q$ phases in the magnetic field,
including the topologically trivial one with $N_{\rm m}=0$. 
We find that $\chi_{\rm sc}$ is drastically reduced at the topological transitions where $N_{\rm m}$ becomes zero, and it vanishes at the transitions to the other $1Q$, 
$2Q$, or PM phases either continuously or discontinuously. 

Let us closely look at the phase transitions under the magnetic field at finite temperature, focusing on the case of $T=0.45$ in ${\bf h}||[110]$, 
where the intermediate $2Q$ phase appears in addition to the multiple $3Q$ phases with different $N_{\rm m}$.
Figures~\ref{fig07}(a)--\ref{fig07}(c) show $h$ dependences of $m_\eta$, $C$, 
the field derivative of the magnetization $dm/dh$, $N_{\rm m}$, and $\chi_{\rm sc}$.
When $h$ is applied, the spiral component $m_3$, which is perpendicular to the magnetic field (${\bf Q}_3 \perp {\bf h}$), 
deviates from the other $3Q$ components ($m_1=m_2$) due to the symmetry breaking by $h$ and decreases as increasing $h$, as shown in Fig.~\ref{fig07}(a). 
With further increasing $h$, $m_3$ continuously vanishes at $h\simeq 0.305$ as $m_3 \propto | h-h_c |^{1/2}$, and the system turns into the $2Q$ phase. 
Due to the disappearance of one of the primary order parameters, the secondary order parameters $m_{4\text{--}7}$ also vanish as $m_{4\text{--}7} \propto |h-h_c|^{1/2}$. 
At the transition, both $C$ and $dm/dh$ show a jump, as shown in Fig.~\ref{fig07}(b).

In addition to this magnetic phase transition, the system undergoes successive topological transitions within the $3Q$ phase at $h\simeq 0.238$ and $0.270$, 
where $N_{\rm m}$ changes stepwise as $N_{\rm m}=8$ to $6$ and to $4$ as shown in Fig.~\ref{fig07}(c).
They are associated with annihilations of a pair of monopole and antimonopole per MUC.
Interestingly, these transitions are not accompanied by any anomalies in $C$ and  $dm/dh$ [green arrows in Fig.~\ref{fig07}(b)], 
in stark contrast to the magnetic phase transition at $h\simeq 0.305$ where the topology also changes as $N_{\rm m}$ vanishes.
We also note that $\chi_{\rm sc}$ does not show any anomalies at these topological transitions, 
while it decreases rapidly along with the decrease of $N_{\rm m}$ after showing a broad maximum in the $N_{\rm m}=8$ phase, as shown in Fig.~\ref{fig07}(c).
Thus, the topological transitions with changes of $N_{\rm m}$, when they are not accompany by the magnetic phase transition, 
are ``hidden'' transitions that do not show any anomalies in the thermodynamic quantities. 
We note that similar hidden transitions were found in the previous study of the ground state~\cite{Okumura2020}, 
but their features, especially at finite temperature, have not been analyzed because of the high computational cost and less accuracy in the numerical simulations.

Through the magnetic phase transition from $3Q$ to $2Q$ at $h\simeq 0.305$, the spin texture changes from the three- to two-dimensional one.
Figure~\ref{fig07}(d) shows the spin configuration and the positions of monopoles and antimonopoles in the $3Q$ phase close to the critical field.
The spin structure is three-dimensional, but the modulation in the direction of ${\bf Q}_3$ ($z$ direction) is weak 
as $m_3$ is small compared to $m_1$ and $m_2$ as shown in the inset.
When entering into the $2Q$ phase, the modulation in the $z$ direction is completely eliminated as $m_3$ becomes zero, and the spin texture becomes two-dimensional.
Figures~\ref{fig07}(e)--\ref{fig07}(g) show the spin configurations on a $(001)$ slice in the $2Q$ phase near the transition.
In this region, we find a vortex-like texture. 
To identify vorticies and antivorticies,
we compute the vorticity defined by the sum of the four relative angles ($\in (-\pi,\pi]$) between the neighboring spins 
surrounding each square plaquette after projecting onto the $(001)$ plane: 
It takes $(-)2\pi$ for the (anti)vortex. 
As a result, we reveal that the $2Q$ state for $0.305 \lesssim h \lesssim 0.365$ consists of two vortex--antivortex pairs, 
whose cores are indicated by the orange and blue plaquettes in Figs.~\ref{fig07}(e) and \ref{fig07}(f).
We show that the vortices and antivortices disappear at $h \simeq 0.365$, 
suggesting another topological transition. 
Again $C$ and $dm/dh$ do not show any anomalies, while both shows a broad hump 
around the topological transition, as shown by the blue arrow in Fig.~\ref{fig07}(b).

%%%%%%%%%%%%%%%%%%%%%%%%%%%%%%%%%%%%
\subsubsection{$p=0.5$}\label{sec:FTp0.5}
%%%%%%%%%%%%%%%%%%%%%%%%%%%%%%%%%%%%
\begin{figure}[bhtp]
  \centering
  \includegraphics[trim=0 10 0 0, clip,width=\columnwidth]{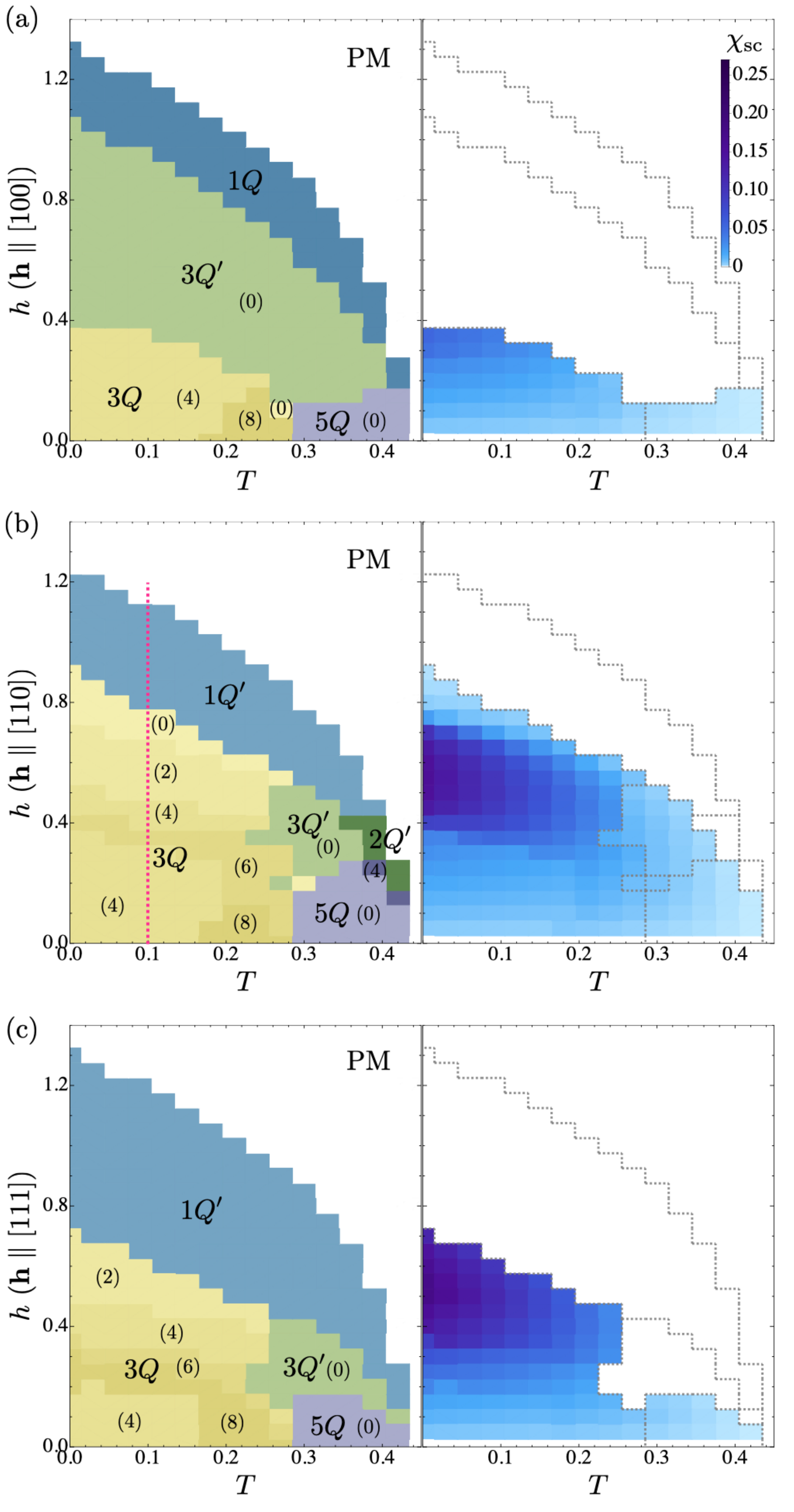}
  \caption{
  Magnetic field--temperature phase diagrams (left)
  and  $h$--$T$ maps of the spin scalar chirality $\chi_{{\rm sc}}$ (right) for 
  (a) ${\bf h}\parallel [100]$,
  (b) ${\bf h}\parallel [110]$, and
  (c) ${\bf h}\parallel [111]$
  at $p=0.5$.
  The magenta dashed line in the left panel of (b) indicates the parameter range for Fig.~\ref{fig10}. 
  The other notations are common to those in Fig.~\ref{fig05}.
  }
  \label{fig08}
\end{figure}
%%%%%%%%%%%%%%%%%%%%%%%%%%%%%%%%%%%%
\begin{figure}[bhtp]
  \centering
  \includegraphics[trim=0 0 0 0, clip,width=\columnwidth]{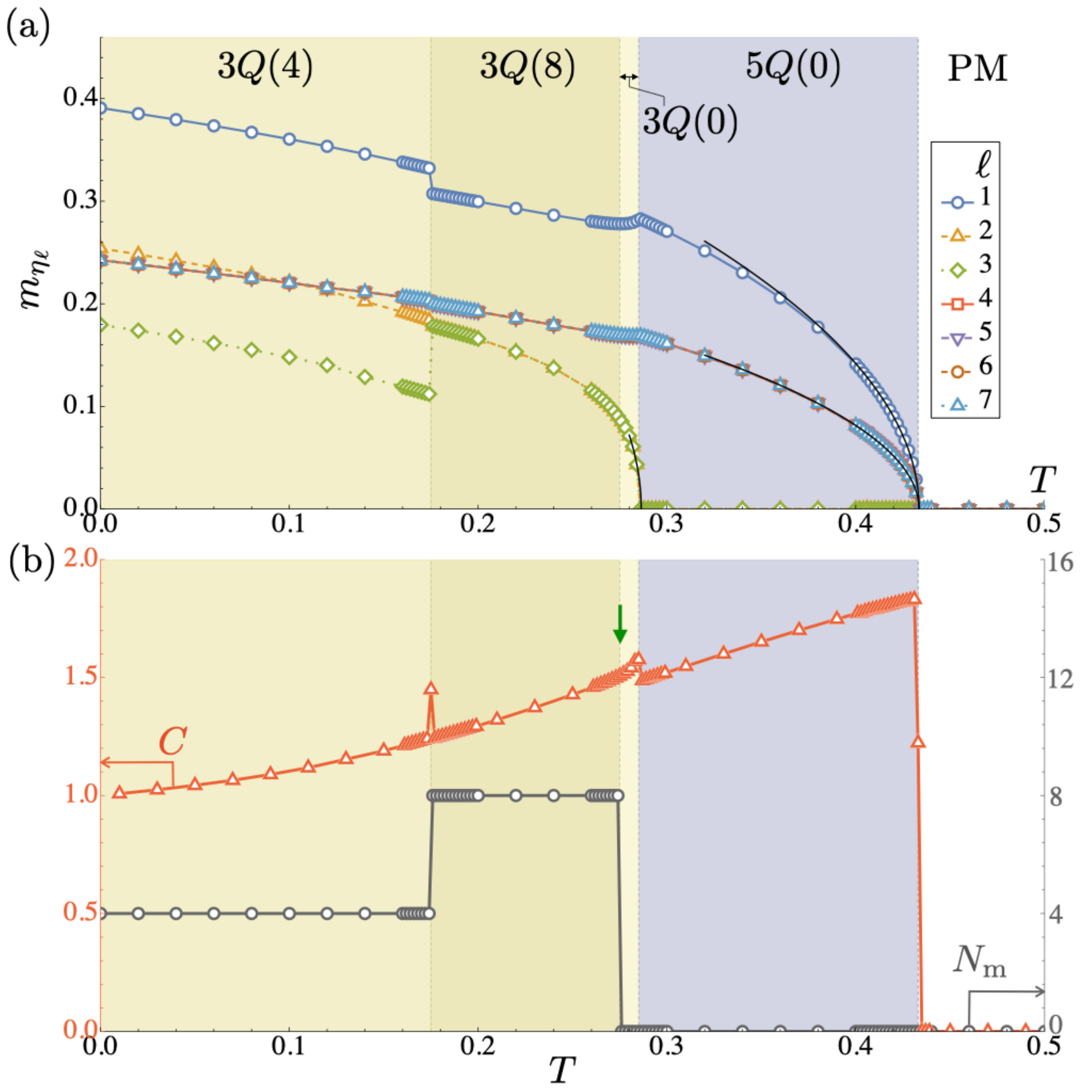}
  \caption{
  Temperature dependences of (a) $m_{\eta_\ell}$ and
  (b) $C$ and $N_{\rm m}$ at $h=0$ and $p=0.5$.
  The black solid lines in (a) indicate functions proportional to $|T-T_c|^{1/2}$.
  }
  \label{fig09}
\end{figure}
%%%%%%%%%%%%%%%%%%%%%%%%%%%%%%%%%%%%

Next, we show the results for $p=0.5$.
Figure~\ref{fig08} summarizes the phase diagrams and $\chi_{{\rm sc}}$ on the $h$--$T$ plane.
At $h=0$, the ground state is the $3Q$ HL with two pairs of monopoles and antimonopoles ($N_{\rm m}=4$); see Fig.~\ref{fig03}(b).
With increasing $T$, the system undergoes successive transitions before entering into the PM state above $T\simeq 0.433$. 
The temperature dependences of $m_\eta$ and $C$ are shown in Fig.~\ref{fig09}. 
At $T\simeq 0.175$, the system shows a first-order phase transition to another $3Q$ HL with $N_{\rm m}=8$, 
where $m_\eta$ show discontinuous changes: 
The primary $3Q$ components change from $m_1>m_2>m_3$ to $m_1>m_2=m_3$, as shown in Fig.~\ref{fig09}(a). 
At the same time, $C$ shows a delta function like anomaly, as shown in Fig.~\ref{fig09}(b). 
Thus, this is a discontinuous phase transition between the $3Q$ HLs with different magnetic and topological properties. 
In contrast, 
the next transition at $T\simeq 0.275$, where $N_{\rm m}$ changes from $8$ to $0$ by annihilation of four pairs of monopoles and antimonopoles, is a hidden topological transition: 
$C$ as well as $m_\eta$ does not show any anomalies as indicated by the green arrow in Fig.~\ref{fig09}(b), similar to those with $N_{\rm m}=8\to 6\to 4$ at $p=0.4$ in the previous section.

When $T$ is further increased, a second-order phase transition occurs from $3Q$ to $5Q$ at $T \simeq 0.285$. 
At this transition, two of $m_{1\text{--}3}$ vanish as $m_\eta \propto |T-T_c|^{1/2}$, leaving one nonzero component of $3Q$ in addition to the four $4Q$ ones. 
While the nonzero magnetic components are common, the $5Q$ state is topologically different from that found in the ground state in Fig.~\ref{fig02}, as the present one have no monopoles ($N_{\rm m}=0$). 
Thus, this continuous transition is purely magnetic, taking place between the topologically-trivial phases. 
We note that all $m_{4\text{--}7}$ show the criticality of $|T-T_c|^{1/2}$ 
at the transition to the high-$T$ PM state at $T\simeq 0.433$,
indicating that the $4Q$ components are not secondary but primary order parameters, unlike those in the $3Q$ phase in Fig.~\ref{fig06}.

In an applied magnetic field, the system undergoes complex successive transitions, as shown in Fig.~\ref{fig08}. 
In particular, the $3Q$ phase experiences multiple changes of $N_{\rm m}$ at low and intermediate temperatures.
Similar to the case of $p=0.4$, $\chi_{\rm sc}$ has nonzero values in all the $3Q$ phases in the magnetic field, as shown in Fig.~\ref{fig08}(b).
While increasing $h$ or $T$, we find another $3Q$ phase dubbed $3Q'$ in all the field directions before going to the $1Q$ or $1Q'$ phase. 
Interestingly, in the $3Q'$ phase, $\chi_{\rm sc}$ is zero for ${\bf h} \parallel [100]$ and $[111]$, but nonzero for ${\bf h} \parallel [110]$.
This peculiar behavior is understood as follows. 
The spin configuration is similar to the one in the $3Q'$ phase found in the ground state in Sec.~\ref{sec:GSpB}, 
which is two-dimensional due to the coplanar arrangement of the three ordering wave numbers. 
In this situation, $\chi_{{\rm sc}} =0$ when the magnetic field is parallel to the two-dimensional plane, 
because $\sum_{{\bf r}_0} {\boldsymbol \chi}_{{\bf r}_0}$ in Eq.~\eqref{eq:chi_sc} becomes perpendicular to the plane.
This condition holds for ${\bf h}\parallel [100]$ and ${\bf h}\parallel [111]$, but not for ${\bf h}\parallel [110]$.
For example, 
$\{1,4,7\}$ is a set of $\eta$ for nonzero $m_\eta$ commonly seen in all the field directions, and its corresponding plane is $(01\bar{1})$ 
that is perpendicular to $[100]$ and $[111]$ but not to $[110]$. 
In addition, we note that $\chi_{\rm sc}$ is nonzero in the $5Q$ phase under the magnetic field despite being topologically trivial, 
but is zero in the $2Q'$ phase appearing at finite $h$ and $T$ as the $2Q$ phase in the case of $p=0.4$.

%%%%%%%%%%%%%%%%%%%%%%%%%%%%%%%%%%%%
\begin{figure}[bhtp]
  \centering
  \includegraphics[trim=0 20 0 0, clip,width=\columnwidth]{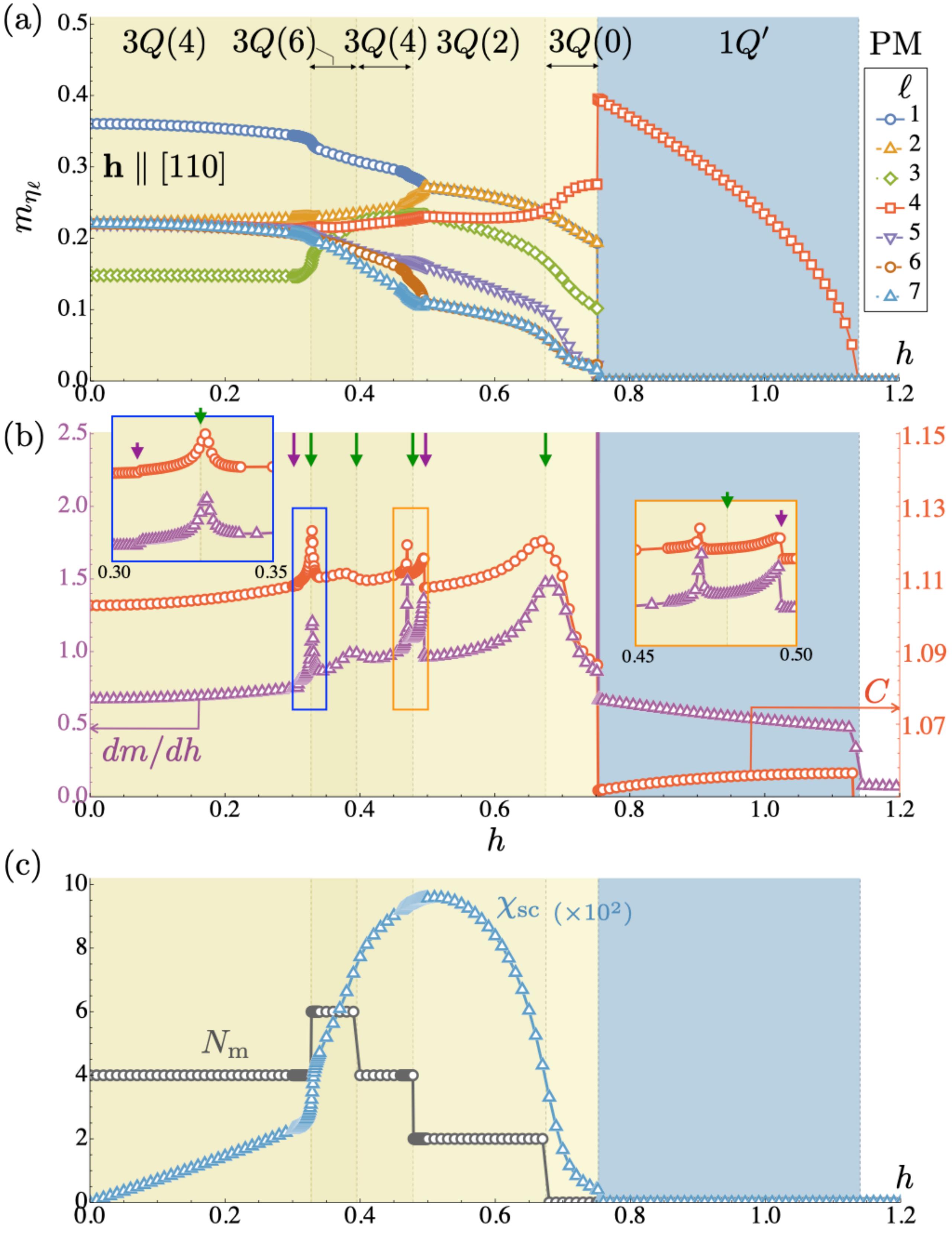}
  \caption{
  Magnetic field dependences of (a) $m_{\eta_\ell}$, 
  (b) $C$ and $dm/dh$, and (c) $N_{\rm m}$ and $\chi_{\rm sc}$. 
  The calculations are done for ${\bf h}\parallel [110]$ at $T=0.1$ and $p=0.5$; see the magenta dashed line in the left panel of Fig.~\ref{fig08}(b). 
  The insets in (b) are magnified views of the blue- and orange-boxed areas in the main panel.
  The data for $\chi_{\rm sc}$ in (c) are multiplied by a factor of $10^2$ for better visibility.
  The green and purple arrows in (b) indicate the hidden topological transitions and the continuous magnetic phase transitions, respectively.
  }
  \label{fig10}
\end{figure}
%%%%%%%%%%%%%%%%%%%%%%%%%%%%%%%%%%%%

Let us discuss the magnetic and topological transitions in the $3Q$ phase, 
by taking the case of $T=0.1$ in ${\bf h}||[110]$ as an example [magenta dashed line in Fig.~\ref{fig08}(b)].
Figure~\ref{fig10} shows the magnetic field dependences of $m_\eta$, $C$, $dm/dh$, $N_{\rm m}$, and $\chi_{\rm sc}$.
When $h$ is applied, $m_\eta$ show a complex field dependence; 
from the anomalies in $m_\eta$, we identify two magnetic phase transitions within the $3Q$ phase at $h\simeq 0.309$ and $h\simeq 0.495$. 
These transitions are of second order and purely magnetic without a change of $N_{\rm m}$; 
$C$ and $dm/dh$ show a jump, as shown by the purple arrows in Fig.~\ref{fig10}(b). 
In addition to these magnetic phase transitions, we find four topological transitions associated with the changes of $N_{\rm m}$ as $N_{\rm m} = 4 \to 6 \to 4 \to 2 \to 0$. 
Similar to the case of $p=0.4$ in Fig.~\ref{fig07}(b), all these transitions are hidden with no anomalies in $C$ and $dm/dh$, as shown by the green arrows in Fig.~\ref{fig10}(b). 
It is, however, worth noting that they exhibit clear humps in both $C$ and $dm/dh$ except for the one at $h\simeq 0.479$, in contrast to the $p=0.4$ case. 
As plotted in Fig.~\ref{fig10}(c), $\chi_{\rm sc}$ shows a sharp rise at the transition with $N_{\rm m}=4\to 6$. 
After the increase through the topological transitions with $N_{\rm m}=6\to 4\to 2$ and showing a peak in the $N_{\rm m}=2$ phase, 
$\chi_{\rm sc}$ rapidly decreases at the transition with $N_{\rm m}=2\to 0$, and goes to zero discontinuously at the magnetic phase transition to the $1Q'$ phase.

%%%%%%%%%%%%%%%%%%%%%%%%%%%%%%%%%%%%
\subsubsection{$p=0.6$}\label{sec:FTp0.6}
%%%%%%%%%%%%%%%%%%%%%%%%%%%%%%%%%%%%
\begin{figure}[bhtp]
  \centering
  \includegraphics[trim=0 10 0 0, clip,width=\columnwidth]{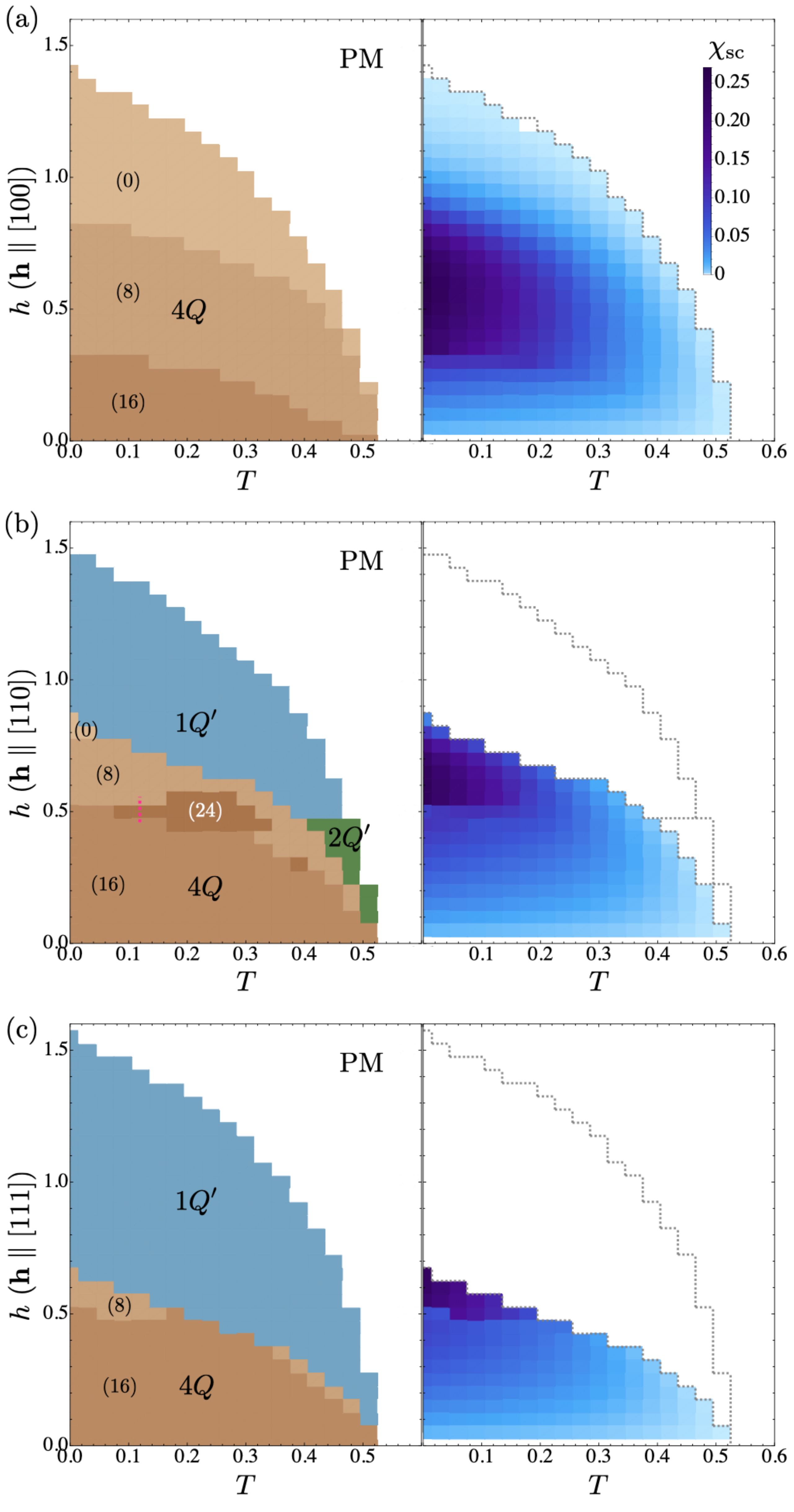}
  \caption{
  Magnetic field--temperature phase diagrams (left)
  and  $h$--$T$ maps of spin scalar chirality $\chi_{{\rm sc}}$ (right) for 
  (a) ${\bf h}\parallel [100]$,
  (b) ${\bf h}\parallel [110]$, and
  (c) ${\bf h}\parallel [111]$
  at $p=0.6$.
  The magenta dashed line in the left panel of (b) indicates the parameter range for Fig.~\ref{fig12}. 
  The other notations are common to those in Fig.~\ref{fig05}.
  }
  \label{fig11}
\end{figure}
%%%%%%%%%%%%%%%%%%%%%%%%%%%%%%%%%%%%
\begin{figure}[bhtp]
  \centering
  \includegraphics[trim=0 28 0 0, clip,width=\columnwidth]{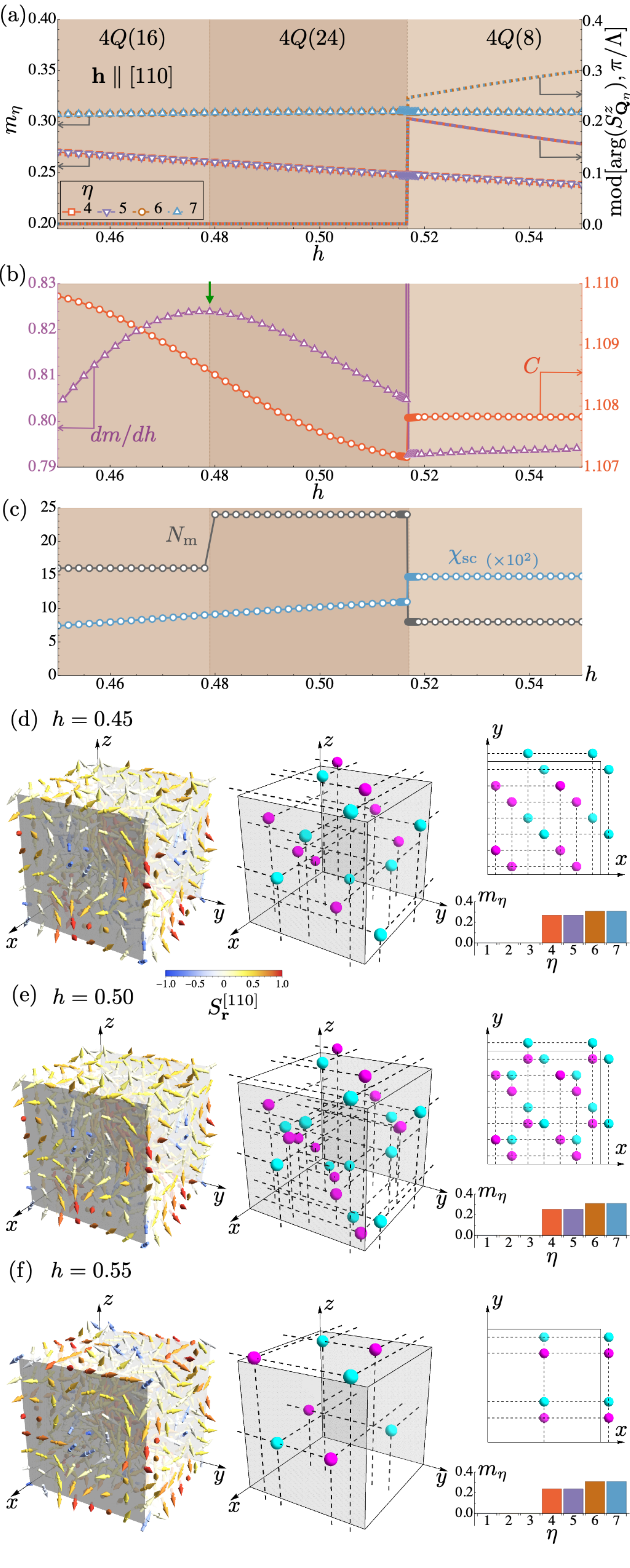}
  \caption{
   Magnetic field dependences of (a) $m_\eta$,  modulo $\pi/\Lambda$ of the argument of $S^z_{{\bf Q}_\eta}$ 
  (${\rm mod}[\arg (S^z_{{\bf Q}_\eta} ), \pi/\Lambda] $),
   (b) $C$ and $dm/dh$, and (c) $N_{\rm m}$ and $\chi_{\rm sc}$ for ${\bf h}\parallel [110]$ at $T=0.12$ and $p=0.6$. 
  The green arrow in (b) indicates the hidden topological transition.
  Spin configurations and positions of monopoles and antimonopoles at (d) $h=0.45$, (e) $h=0.50$, and (f) $h=0.55$.
  The insets show distributions of $m_\eta$.
  }
  \label{fig12}
\end{figure}
%%%%%%%%%%%%%%%%%%%%%%%%%%%%%%%%%%%%

Finally, we show the results for $p=0.6$, which is on the $4Q$ side in the competing region; see Figs.~\ref{fig02} and \ref{fig04}.
Figure~\ref{fig11} shows the $h$--$T$ phase diagrams and $\chi_{{\rm sc}}$.
At $h=0$, the ground state is the $4Q$ HL with $N_{\rm m}=16$, which remains stable until the phase transition to the PM phase at $T\simeq 0.52$.
In an applied magnetic field,
the system undergoes a direct transition from the $4Q$ phase to the PM phase for ${\bf h}\parallel [100]$, but it shows an additional phase transition
to the $1Q'$ phase for ${\bf h}\parallel [110]$ and $[111]$. 
In addition, the $2Q'$ phase appears at finite $T$ under ${\bf h}\parallel [110]$, similar to the case of $p=0.5$ in Fig.~\ref{fig08}(b).
In the $4Q$ phase, $N_{\rm m}$ varies with $T$ and $h$. 
$\chi_{\rm sc}$ is nonzero in the entire region of the $4Q$ phase, 
and becomes large in the $N_{\rm m}=8$ phase at low temperature.
The largest value of $N_{\rm m}=24$ is found at finite temperature under ${\bf h} \parallel [110]$.

Let us discuss the transitions including the $4Q$ HL with the largest $N_{\rm m} = 24$, 
by taking the case of $T=0.12$ in ${\bf h}||[110]$ [magenta dashed line in Fig.~\ref{fig11}(b)]. 
Figure~\ref{fig12} shows the magnetic field dependences of $m_\eta$, $C$, $dm/dh$, $N_{\rm m}$, and $\chi_{\rm sc}$.
In this range of $h$, $N_{\rm m}$ changes from $16$ to $24$ at $h\simeq 0.479$ and from $24$ to $8$ at $h\simeq 0.517$. 
Let us first discuss the transition at $h\simeq 0.479$. 
We find that this is a hidden topological transition with no anomalies in $m_\eta$, $C$, and $dm/dh$, except for a hump in $C$ as indicated by the green arrow in Fig.~\ref{fig12}(b). 
Figures~\ref{fig12}(d) and \ref{fig12}(e) show the configurations of spins and monopoles in the $N_{\rm m}=16$ and $N_{\rm m}=24$ phases, respectively. 
The intriguing aspect of this topological transition is that the monopoles and antimonopoles are pair created for the increase of $h$ without a magnetic phase transition. 
Although there are several examples of pair annihilation for the increase of the magnetic field~\cite{Okumura2020,Okumura2020b,Shimizu2021a,Okumura2022}, 
this is the first example of pair creation in the $4Q$ HLs to the best of our knowledge.

Next let us discuss the transition at $h \simeq 0.517$. 
This is a first-order magnetic phase transition with small jumps of $m_\eta$. 
Through the discontinuous change of the spin configuration, $N_{\rm m}$ also changes from $24$ to $8$; 
see Figs.~\ref{fig12}(e) and \ref{fig12}(f). 
Interestingly, we find that the phase transition is accompanied by 
changes in the complex phases of the Fourier components $S^\mu_{{\bf Q}_\eta}$, as plotted in Fig.~\ref{fig12}(a); 
we here plot $\mod [\arg(S^z_{{\bf Q}_{4\text{--}7}}), \pi/\Lambda]$, where ${\rm mod}[x,y] \equiv x - y\lfloor x/y \rfloor$, 
as the complex phase has arbitrariness of multiples of $\pi/\Lambda$ corresponding to spatial translation.
Note that the importance of such phase degree of freedom 
in topological spin textures has also been pointed out in Refs.~\cite{Hayami2021c,Shimizu2022}. 

%%%%%%%%%%%%%%%%%%%%%%%%%%%%%%%%%%%%
\section{Summary}\label{sec:summary}
%%%%%%%%%%%%%%%%%%%%%%%%%%%%%%%%%%%%

In summary, we have theoretically investigated topological transitions 
driven by external magnetic fields and temperature in emergent magnetic monopole lattices.
Motivated by the recent experimental discovery of the monopole lattices in magnets,
we have proposed a spin model that can stabilize both cubic $3Q$ and tetrahedral $4Q$ HLs and studied the topological nature of the magnetic and thermodynamic properties.
The ground-state phase diagrams~(Figs.~\ref{fig02} and \ref{fig04}) and the magnetic field--temperature phase diagrams~(Figs.~\ref{fig05}, \ref{fig08}, and \ref{fig11}) 
have been obtained precisely using the recently developed exact steepest descent method
that is crucial to identify the topological transitions in the thermodynamic limit.
Through the comprehensive analyses, we have found a variety of hidden topological transitions which do not show anomalies in the macroscopic physical quantities 
such as the specific heat, 
the magnetization, and the net spin scalar chirality, whereas the other magnetic transitions with topological changes exhibit critical behaviors like in the conventional phase transitions. 
Some of the hidden topological transitions show humps in the magnetic field and temperature dependences of the macroscopic quantities like crossovers, but the others do not --- the latter are hardly visible in macroscopic measurements. 
These findings indicate that one needs to be extremely careful to identify such topological transitions associated with the emergent magnetic monopoles.

While our model is constructed by simply integrating the models for the $3Q$ and $4Q$ HLs, it reproduces well some aspects of the experimental results for \ce{MnSi_{1-$x$}Ge_{x}} by regarding $x$ as the mixing ratio $p$ of the magnetic interactions;
the relative stability of the two HLs is sensitive to $p$ but almost insensitive to the magnetic field $h$ 
(Figs.~\ref{fig02} and \ref{fig04}).
Furthermore, our model predicts unprecedented $5Q$ states both with and without magnetic monopoles in the competing region between the $3Q$ and $4Q$ HLs, and a different type of topological transition associated with pair annihilation of two-dimensional magnetic vortices and antivortices in the $2Q$ state appearing only at finite temperature under a magnetic field. 
For further quantitative comparison including these new findings, 
however, experimental studies of the single crystals are indispensable.
Furthermore, it would be important to elaborate a more sophisticated model by taking into account, e.g., 
the single-ion magnetic anisotropy, short-range magnetic interactions, 
magnetic field dependences of the coupling constants as well as the lattice structures, 
and randomness by chemical substitutions.
We hope the present results stimulate such future experimental and theoretical studies for developing the emergent electromagnetism.
\vspace{0.5cm}

\begin{acknowledgments}
The authors would like to thank Y. Fujishiro, S. Hayami, 
N. Kanazawa, Yusuke Kato, A. Miyake, 
S. Okumura, K. Shimizu, 
and M. Tokunaga for fruitful discussions.
Y. K. would like to thank K. Inui 
for his assistance in using 
the JAX-based library, Optax.
This work was supported by Japan Society for the Promotion of Science (JSPS) KAKENHI Grant Nos.~JP19H05825 and JP22K03509,
and JST CREST Grant No.~JPMJCR18T2.
\end{acknowledgments}

%%%%%%%%%%%%%%%%%%%%%%%
%\renewcommand{\thesection}{\Alph{section}}
%\renewcommand{\thesubsection}{\Alph{section}.\arabic{subsection}}
%\renewcommand{\thesubsubsection}{\Alph{section}.\arabic{subsection}.\arabic{subsubsection}}
%\renewcommand{\thefigure}{S\arabic{figure}}
%\renewcommand{\theequation}{S\arabic{equation}}
%\renewcommand{\thetable}{S\Roman{table}}
%\baselineskip=6mm
%%%%%%%%%%%%%%%%%%%%%%%
%\appendix
%\section{}
%%%%%%%%%%%%%%%%%%%%%%%

\bibliography{draft} 

\end{document}